\documentclass[aps,twocolumn,preprintnumbers,amsmath,amssymb,pra,letterpaper,superscriptaddress]{revtex4-1}
\usepackage{multirow}

\usepackage{amsmath,amssymb}
\usepackage{graphicx}
\begin{document}

\author{Nicholas R. Hutzler}
\email{nick@cua.harvard.edu}
\affiliation{Harvard University Department of Physics, 17 Oxford Street, Cambridge, MA 02138 USA}
\author{Hsin-I Lu}
\email{hsin-i@cua.harvard.edu}
\affiliation{Harvard University Department of Physics, 17 Oxford Street, Cambridge, MA 02138 USA}
\affiliation{Harvard University School of Engineering and Applied Sciences, 29 Oxford Street, Cambridge, MA 02138 USA}
\author{John M. Doyle}
\email{doyle@physics.harvard.edu}
\affiliation{Harvard University Department of Physics, 17 Oxford Street, Cambridge, MA 02138 USA}
\title{The Buffer Gas Beam: An Intense, Cold, and Slow Source for Atoms and Molecules}
\maketitle

\newcommand{\kn}{Kn}
\newcommand{\re}{Re}
\newcommand{\ma}{Ma}
\newcommand{\be}{\begin{equation}}
\newcommand{\ee}{\end{equation}}
\newcommand{\bea}{\begin{eqnarray}}
\newcommand{\eea}{\end{eqnarray}}

\tableofcontents

\section{Introduction}\label{sec:introduction}

Beams of atoms and molecules are stalwart tools for spectroscopy and studies of collisional processes\cite{Scoles1988,Ramsey1985}.  The supersonic expansion technique can create cold beams of many species of atoms and molecules.  However, the resulting beam is typically moving at a speed of 300--600 m s$^{-1}$ in the lab frame, and for a large class of species has insufficient flux (i.e. brightness) for important applications.  In contrast, buffer gas beams\cite{Maxwell2005,Campbell2009} can be a superior method in many cases, producing cold and relatively slow molecules in the lab frame with high brightness and great versatility. There are basic differences between supersonic and buffer gas cooled beams regarding particular technological advantages and constraints.  At present, it is clear that not all of the possible variations on the buffer gas method have been studied.  In this review, we will present a survey of the current state of the art in buffer gas beams, and explore some of the possible future directions that these new methods might take.

Compared to supersonic expansion, the buffer gas cooled beam method has a fundamentally different approach to cooling molecules into the kelvin regime.  The production of cold molecules (starting from hot molecules) is achieved by initially mixing two gases in a cold cell (with dimensions of typically a few cm).  The two gases are the hot ``species of interest'' molecules (introduced at an initial temperature $T_0$ typically between 300--10,000 K)  and cold, inert ``buffer'' gas atoms (cooled to 2-20 K by the cold cell).  The buffer gas in the cell is kept at a specifically tuned atom number density, typically $n = 10^{14-17}$ cm$^{-3}$, which is low enough to prevent simple three body collision cluster formation involving the target molecule, yet high enough to provide enough collisions for thermalization before the molecules touch the walls of the cold cell.  A beam of cold molecules can be formed when the buffer gas and target molecules escape the cell through a few-millimeter-sized orifice, or a more complicated exit structure, into a high vacuum region as shown in Figure \ref{fig:buffergasbeam} .  For certain buffer gas densities, the buffer gas aids in the extraction of molecules into the beam via a process called ``hydrodynamic enhancement.''\cite{Patterson2007}.  Finally, in the case where the mass of the molecule is higher than that of the buffer gas atom, there is a velocity-induced angular narrowing of the molecular beam, which increases the on-axis beam intensity\cite{Patterson2007}.  Although such enhancement has long been recognized in room-temperature beams\cite{Anderson1967}, it is seldom employed because it requires intermediate Reynolds numbers, in conflict with the high Reynolds numbers necessary for full supersonic cooling of molecules in the beam.  In buffer gas cooled beams, on the other hand, the intermediate Reynolds number regime is often ideal for cooling, and allows this technique to take full advantage of the intensity enhancement from angular narrowing.

With cryogenic cooling, high gas densities are not needed in the buffer cell to cool into the kelvin regime.  In the case of supersonic beams, the high gas densities required in the source can be undesirable.  In the case of buffer gas beams, the cryogenic environment and relatively low flow of buffer gas into the high vacuum beam region allows for 100$\%$ duty cycle (continuous) beam operation, without relying on external vacuum pumps. Rather, internal cryopumping provides excellent vacuum in the beam region. This combination of characteristics has led to high-brightness cold molecule sources (see Tables \ref{table:beamcomparison} and \ref{table:beamproperties}), for both chemically reactive (e.g. pulsed cold ThO, producing $3\times 10^{11}$ ground state molecules per steradian during a few ms long pulse\cite{Hutzler2011}) and stable molecules (e.g. continuous cold O$_2$, producing $\approx3\times 10^{13}$ cold molecules per second\cite{Patterson2007,OxygenFlux}).

\begin{table*}
\begin{tabular}{llll}
Method 			& Species & Intensity [s$^{-1}]$ & Velocity [m s$^{-1}$] \\ \hline
\multicolumn{4}{l}{\emph{Chemically reactive polar molecules}} \\
Buffer gas 	& ThO\cite{Hutzler2011}	&	$3\times 10^{13}$	sr$^{-1}$				& 170	 \\
Buffer gas 	& SrF\cite{Barry2011}		&	$1.7\times 10^{12}$	sr$^{-1}$			& 140	 \\
Buffer gas 	& CaH\cite{Lu2011}			&	$5\times 10^9$ sr$^{-1}$					& 40	 \\
Supersonic				& YbF\cite{Tarbutt2002}	&	$1.4\times 10^{10}$ sr$^{-1}$	&	290 \\
Supersonic				& BaF\cite{Rahmlow2010}	&	$1.3\times 10^{10}$ sr$^{-1}$	&	500 \\
Effusive					& SrF\cite{Tu2009}			& $5\times10^{11}$ sr$^{-1}$	&	650	\\
Effusive					& ThO\cite{Hutzler2011,Ackermann1973}	& $1\times10^{11}$ sr$^{-1}$ &	540			\\ \hline
\multicolumn{4}{l}{\emph{Stable molecule with significant vapor pressure at 300 K}} \\
Buffer gas 	& ND$_3$\cite{Buuren2009,Sommer2009} 		& $1\times 10^{11}$		& 65 \\
Supersonic & ND$_3$\cite{Bethlem2000,Bethlem2002}	& $1\times 10^8$		& 280 \\
Stark Decelerated	& ND$_3$\cite{Bethlem2000,Bethlem2002}	& $1\times 10^6$			& 13 \\
Effusive 		& ND$_3$\cite{Junglen2004} 							& $2\times 10^{10}$ 	& 40 		\\
\end{tabular}		
\caption{A comparison of supersonic, effusive, and buffer gas cooled beams for selected molecules.  The intensity is the number of molecules per quantum state per second, and the velocity is reported in the forward direction.  If the angular spread of the beam is known, then the intensity per unit solid angle (i.e. brightness) is given.  The ND$_3$ sources are velocity-selected.  For pulsed sources, the intensity is averaged over a time period that includes multiple pulses.  For chemically reactive molecules, buffer gas sources can provide dramatically higher brightness, and with a slower forward velocity.  In addition to brightness and forward velocity, we shall see later in this article that other features of buffer gas beams also compare favorably to supersonic and effusive beams.  See Table \ref{table:beamproperties} for more data about experimentally realized buffer gas cooled beams.}
\label{table:beamcomparison}
\end{table*}

\begin{figure}[htbp]
	\centering
		\includegraphics[width=0.45\textwidth]{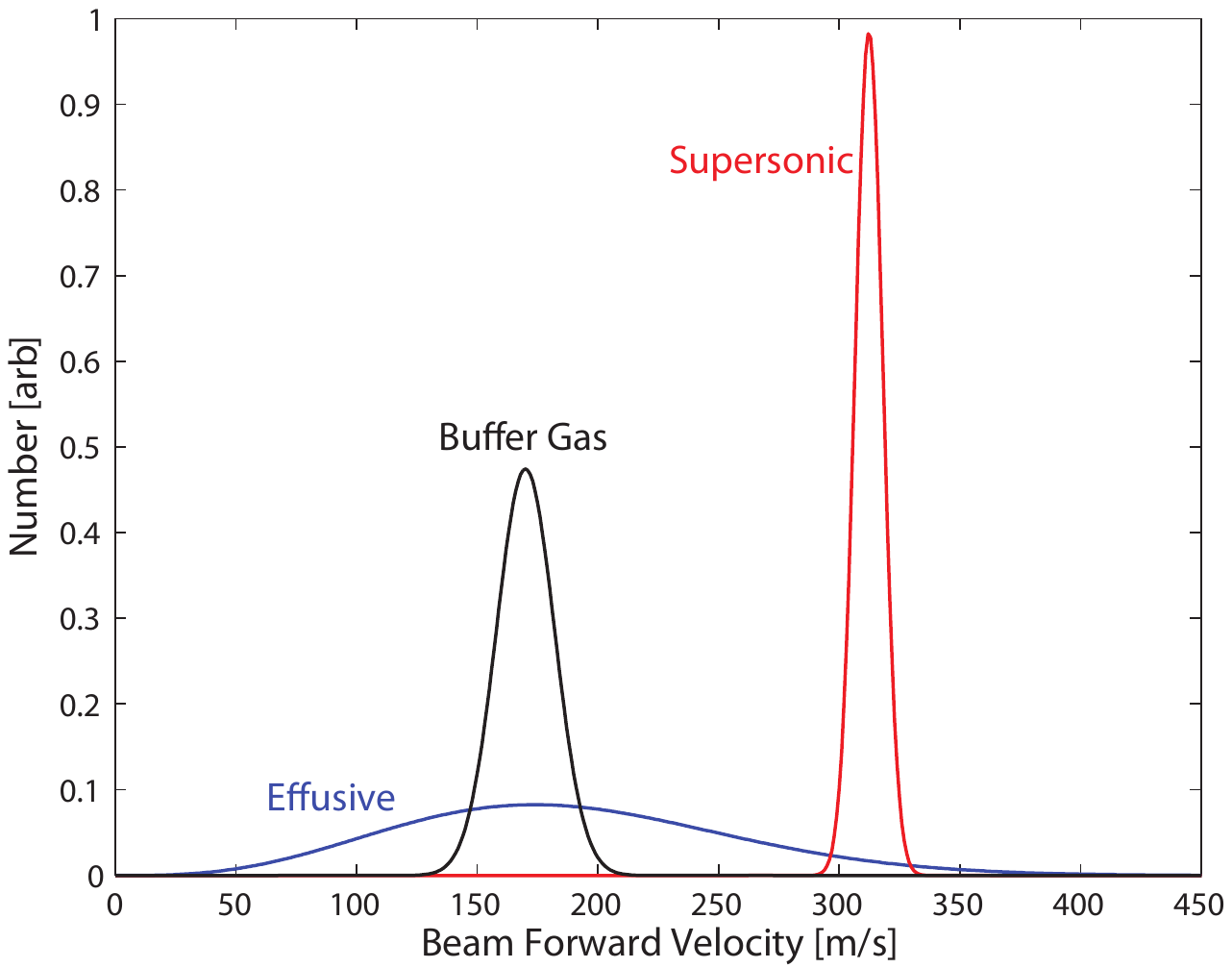}
	\caption{Schematic velocity distributions for selected effusive, supersonic, and buffer gas cooled beam sources.  The buffer gas cooled beam properties are taken from a ThO source with neon buffer gas\cite{Hutzler2011}.  The effusive beam is a simulated room temperature source of a species with mass of 100 amu, and the simulated supersonic source uses room temperature xenon as the carrier gas.  Compared to the effusive beam, the buffer gas beam has a much lower temperature (i.e. smaller velocity spread).  Compared to the supersonic beam, the buffer gas beam has a comparable temperature but lower forward velocity.  Supersonic sources typically have much higher average forward velocity than the one presented above ($\approx 600$ m s$^{-1}$ for room temperature argon, or $\approx 1800$ m s$^{-1}$ for room temperature helium),\cite{Scoles1988} and effusive sources for many species (like ThO\cite{Hutzler2011,Ackermann1973}) would require oven temperatures of $>1000$ K, making the distribution much wider with a much higher mean.   The distributions above are normalized; however, the buffer gas cooled source of many species has a considerably higher flux.  See Table \ref{table:beamcomparison} for data about experimentally realized buffer gas cooled beams.}
	\label{fig:veldist_comparison}
\end{figure}

\subsection{Cold atoms and molecules}
In the past two decades, the evaporative cooling of atoms to ultracold temperatures at high phase space density has opened new chapters for physics and led to exciting discoveries, including the realization of Bose-Einstein condensation\cite{Anderson1995,Davis1995}, strongly correlated systems in dilute gases\cite{Chin2004,Zwierlein2005}, and controlled quantum simulation\cite{Simon2011,Greiner2002}.  Meanwhile, the success of cold atom methods and new theory has inspired the vigorous pursuit of molecule cooling.  Molecules are more complicated than atoms and possess two key features not present in atoms: additional internal degrees of freedom, in the form of molecular rotation and vibration, and the possibility (with polar molecules) to exhibit an atomic unit of electric dipole moment in the lab frame, which can lead to systems with long range, anisotropic, and tunable interactions.  The rich internal structure and chemical diversity of molecules could provide platforms for exploring science in diverse fields, ranging from fundamental physics, cold chemistry, molecular physics, and quantum physics\cite{Carr2009}.  We will list just a few of these applications here.
\begin{itemize}
\item Molecules have enhanced sensitivity (as compared to atoms) to violations of fundamental symmetries, such as the possible existence of the electron electric dipole moment\cite{Hudson2011,Khriplovich1997}, and parity-violating nuclear moments\cite{Ginges2004,Dzuba2010}.
\item The internal degrees of freedom of polar molecules have been proposed as qubits for quantum computers\cite{DeMille2002}, and are ideal for storage of quantum information\cite{Rabl2006,Andre2006,Carr2009}.
\item The long-range electric dipole-dipole interaction between polar molecules may give rise to novel quantum systems\cite{Micheli2006,Lahaye2009}.
\item Precision spectroscopy performed on vibrational or hyperfine states of cold molecules can probe the time variation of fundamental constants, such as the electron-to-proton mass ratio and the fine structure constant \cite{DeMille2008_II,Hudson2006,Flambaum2009}.
\item Studies of cold molecular chemistry in the laboratory play an important role in understanding gas-phase chemistry of interstellar clouds, which can be as cold as 10 K \cite{Schnell2009,Hummon2008,Klemperer2006}.
\item Ultracold chemical reactions have been observed at a temperature of a few hundred nK, with reaction rates controllable by external electric fields\cite{Ospelkaus2010,Ni2010}.
\item Molecular collisions in the few partial wave regime reveal the molecular interaction in great detail\cite{CampbellPRL2009,Gilijamse2006,Krems2008}.
\end{itemize}
The most common (but not the only\cite{BufferGasBEC}) way to cool atoms to ultracold temperatures, defined as the temperature (typically $\lesssim$1 mK) where only $s-$wave collisions occur (for bosons), is laser cooling\cite{Metcalf1999}.  This technique relies on continuously scattering photons from atoms to dissipate the atom's motional energy.  Typically $\sim 10^4$ photon scattering events are needed to significantly reduce the kinetic energy of the atom.  Molecules generally lack closed transitions that can easily cycle this many photons because the excited states of the molecules can decay to many vibrational or rotational states.  Because laser cooling molecules is more difficult than it is with atoms (and has only been demonstrated relatively recently\cite{Shuman2009,Shuman2010}), there continue to be broad efforts in developing new molecule cooling methods in order to fully control the internal and motional degrees of freedom of molecules\cite{Doyle2004,Friedrich2009,Carr2009,Bethlem2003}.  Additionally, many of these new proposed methods could work well with laser cooling, in some form.

In general, cooling techniques for molecules can be broadly classified into two types: indirect and direct\cite{Doyle2004,Friedrich2009}.  Indirect cooling relies on assembling two laser-cooled atoms into a bound molecule using photoassociation or magnetoassociation.  The resulting molecules have the same translational temperature as their ultracold parent atoms but are typically in a highly excited vibrational state, which can be transferred (typically by a coherent transfer method, such as STIRAP\cite{Bergmann1998}) to the absolute ground state.  While indirect methods can access ultracold polar molecules with a high phase space density, these molecules are currently limited to a combination of the small subset of atoms which have been laser and evaporatively cooled, such as the alkali and alkaline earth atoms.  Currently, high-density samples of ultracold polar molecules have only been demonstrated with a single species, KRb\cite{Ni2008}.  Since many atoms are not easily amenable to laser cooling, there is a large class of molecules that are the focus of direct cooling; this class includes many of the molecules that are desirable for the applications listed above\cite{Carr2009}.  In this paper we shall focus on beams of molecules that are cooled directly through collisions with cryogenically cooled atoms.

\subsection{Buffer gas cooling and beam production}

\begin{figure}[htbp]
 \centering
  \includegraphics[width=0.45\textwidth]{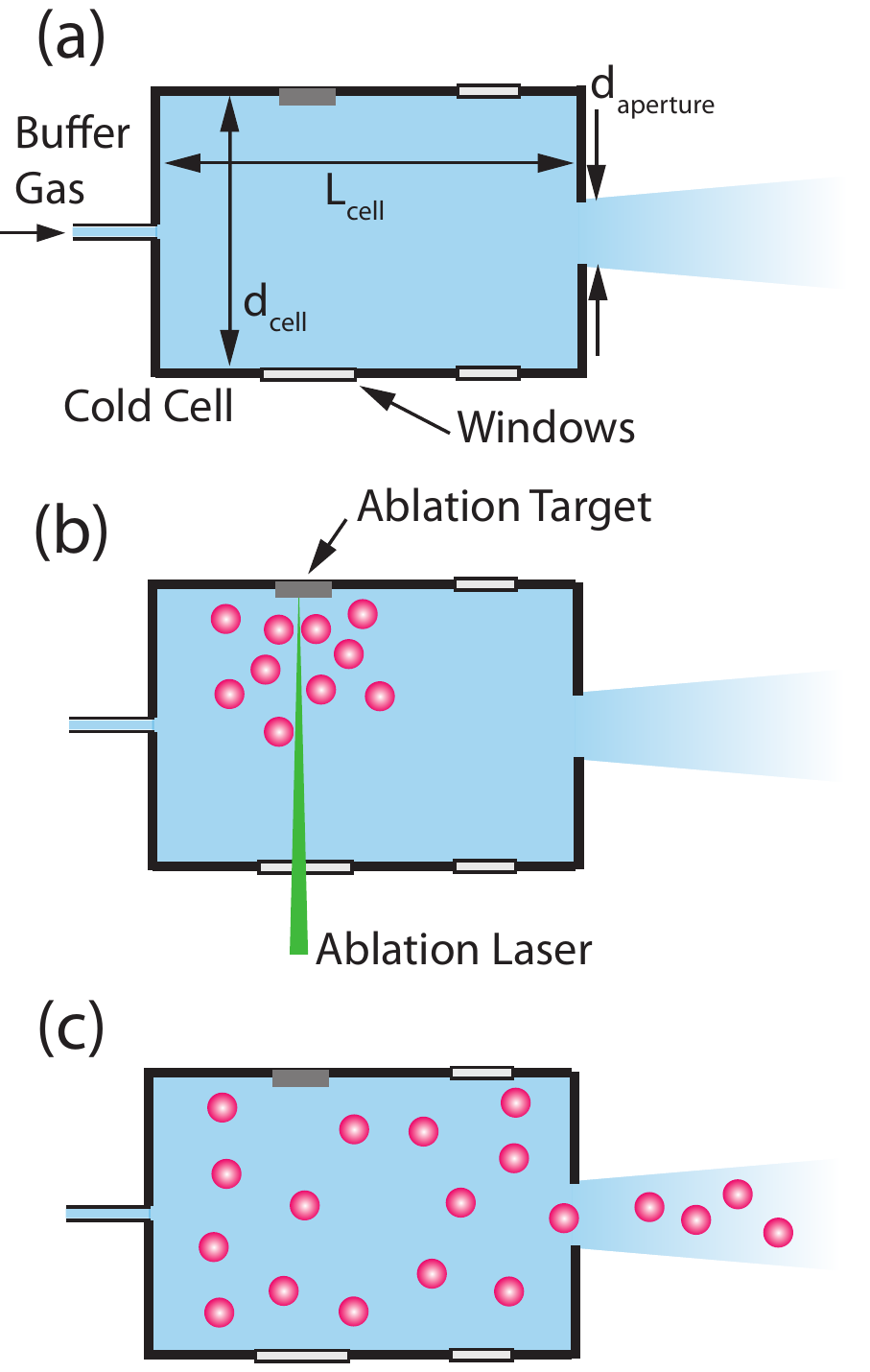}
  \caption{A schematic of a buffer gas beam cell, which is maintained at a temperature of few K. (a) The buffer gas (typically helium or neon) enters the cell through a fill line and exits via a cell aperture on the other side of the cell.  The important physical dimensions (cell length $L_{cell}$, cell diameter $d_{cell}$, and aperture diameter $d_{aperture}$) are indicated. (b) Introduction of species via laser ablation. (c) Molecules thermalize to the buffer gas and form a molecular beam with the flowing buffer gas.}
  \label{fig:buffergasbeam}
\end{figure}
At the heart of the buffer gas beam technique is buffer gas cooling, a direct cooling technique used in the first magnetic trapping of polar molecules over a decade ago\cite{Weinstein1998}.  The buffer gas cooling technique works by dissipating the energy of the species of interest via elastic collisions with cold, inert gas atoms, such as helium or neon. Since this cooling mechanism does not depend on the internal structure of the species (unlike laser cooling), buffer gas cooling can be applied to nearly any atom or small molecule\cite{Campbell2009}, and certain large molecules \cite{Patterson2010II}. Helium maintains a sufficient vapor pressure down to a few hundred mK\cite{Pobell1996,Campbell2009}, and the typical helium-molecule elastic cross section\cite{Campbell2009} of $\sim 10^{-14}$ cm$^{2}$ allows buffer gas cooling and trap loading to be realized with modest cell sizes (few $\times$ few $\times$ few cm$^3$).  Buffer gas cooling (using both helium and neon) has been used to create beams, load magnetic traps, and perform spectroscopy in a cold buffer gas cell\cite{Campbell2009}.  In this paper we shall focus only on one aspect, the creation (and characterization) of buffer gas cooled beams.

As shown in Figure \ref{fig:buffergasbeam}, a buffer gas beam is formed simply by adding an exit aperture to one side of a cold cell filled with a buffer gas.  A constant buffer gas density is maintained by continuously flowing the buffer gas through a fill line on the other side of the cell.  Figure \ref{fig:buffergasbeam}(b) shows laser ablation, one of several methods used to introduce molecules of interest into the cell (see discussion in Section \ref{sec:introductionofspecies}).  After production and injection of molecules in the buffer gas cell, the molecules thermalize with the buffer gas translationally and rotationally, and are carried out of the cell with the flowing buffer gas, forming a molecular beam (Figure \ref{fig:buffergasbeam}(c)).

Cold molecular beams created from a buffer gas cell are made in a few key steps: the introduction of ``hot'' molecules in a ``cold'' buffer gas cell, the thermalization of the molecules with the buffer gas, and the extraction of molecules into a beam.  Each of these steps affects the properties of the resultant beam.   The detailed characteristics (forward velocity, velocity spreads, temperatures, and fluxes) of buffer gas beams in several operating regimes will be described in the next section.  Techniques to manipulate the beam properties, such as guiding, filtering, cooling by isentropic expansion, and advanced cell geometries will also be discussed.  The last part of this paper will discuss the applications of buffer gas beams, including the search for the electron electric dipole moment using ThO\cite{Vutha2010}, the study of cold dipolar collisions with ND$_3$\cite{Sawyer2011}, and the realization of direct laser cooling of molecules with SrF\cite{Shuman2009,Shuman2010}.

\section{Buffer Gas Cooled Beams}

The details of buffer gas cooled beam production and properties will be presented in this section.  In our treatment, we shall assume that the buffer gas is a noble gas, and that the species of interest is seeded in the buffer gas flow with a low fractional concentration, about 1$\%$ or less.  We will often refer to the species of interest as ``the molecule,'' even though buffer gas cooled beams of atoms (e.g. Yb) are also of interest.  As is typically the case, we shall assume that the species of interest is heavier than the buffer gas (though the analysis is easily extended to lighter molecules).

\subsection{Species production, thermalization, diffusion, and extraction}\label{sec:cell}

In this section we present estimates of physical parameters that can be used to support an intuitive understanding of the processes occurring in the buffer gas cell.  These derivations will all be approximate, and will vary depending on geometry, species, introduction method, temperature range, density range, etc.  However, they are typically descriptive within an order of unity.

\subsubsection{Buffer gas flow through the cell}
Consider a buffer gas cell as depicted in Figure \ref{fig:buffergasbeam}.  The cell has a volume of $V_{cell} = A_{cell}\times L_{cell}$, where $L_{cell}$ is the length of the cell interior, and $A_{cell}\approx d_{cell}^2$ is the cross-sectional area (which may be round or square, but has a characteristic length of $d_{cell}$).  $L_{cell}$ is typically a few cm, and $A_{cell}$ is typically a few cm$^2$.  The cell is held at a fixed temperature $T_0$ by a cryogenic refrigerator, typically between 1 K and 20 K.  Buffer gas of mass $m_b$ is introduced into the cell volume by a long, thin tube, or ``fill line''.  The buffer gas exits the cell through an aperture of characteristic length $d_{aperture}$ and area $A_{aperture}$ (for the case where the aperture is a rectangle, $d_{aperture}$ is the shorter dimension).  Typical values are $d_{aperture}=1-5$ mm, and $A_{aperture} = 5-25$ mm$^2$.
A buffer gas flow rate $f_{0,b}$ into the cell can be set with a mass flow controller.  Here the subscript ``$b$'' refers to the buffer gas, and ``0'' refers to conditions in the cell at steady-state, or ``stagnation'' conditions.  The most commonly used unit for flow is the standard cubic centimeter per minute, or SCCM, which equals approximately $4.5\times 10^{17}$ gas atoms per second.  Typical flow rates for buffer gas beams are $f_{0,b}=1-100$ SCCM.  Technical details of cell construction, and construction of a buffer gas cooled beam apparatus in general, are discussed in Section \ref{sec:technicaldetails}.

At steady state, the flow rate out of the cell is given by the molecular conductance of the aperture, $f_{out}=\frac{1}{4}n_{0,b}\bar{v}_{0,b} A_{aperture}$, where
\be\bar{v}_{0,b} = \sqrt{\frac{8k_BT_0}{\pi m_b}} \label{vthermal}\ee
 is the mean thermal velocity of the buffer gas inside the cell (about 140 m s$^{-1}$ for 4 K helium or 17 K neon), $k_B$ is Boltzmann's constant, and $n_{0,b}$ is the stagnation number density of buffer gas atoms.  Therefore, the number density $n_{0,b}$ is set by controlling the flow via the steady-state relationship $f_{out}=f_{0,b}$, or
\be n_{0,b} = \frac{4 f_{0,b}}{A_{aperture}\bar{v}_{0,b}} \label{flowtodensity}.\ee
With typical cell aperture sizes and temperatures, a flow of 1 SCCM corresponds to a stagnation density of about $10^{15}$ cm$^{-3}$, so typical values for the stagnation density are $10^{15}-10^{17}$ cm$^{-3}$.  Note that in the above equation we assume that the flow through the aperture is purely molecular.  At higher number densities the flow can become more fluid-like, and then the flow rate out will change; however, the difference is about a factor of two \cite{Pauly2000}, so the above equation is suitable for approximation.

\subsubsection{Introduction of species}{\label{sec:introductionofspecies}

The species of interest can be introduced into the buffer gas cell through a number of methods \cite{Campbell2009}, including laser ablation, laser-induced atomic desorption (LIAD), beam injection, capillary filling, and discharge etching.  The most commonly used techniques for creation of buffer gas cooled beams are laser ablation and capillary filling, so we will focus here on those two methods.

\emph{Laser ablation.}  In laser ablation (see Figure \ref{fig:buffergasbeam}), a high energy pulsed laser is focused onto a solid precursor target.  After receiving the laser pulse, the solid precursor can eject gas-phase atoms or molecules of the desired species, often along with other detritus.  The actual mechanism for how gas phase species results from the ablation of the solid precursor is not simple, and depends on the relationship between the length of the laser pulse, and the time constants for electronic and lattice heating \cite{Chichkov1996,Olander1990}.  While pulsed ablation has been studied with pulse widths from femtoseconds to milliseconds, the most common ablation laser (used for the majority of experiments in Table \ref{table:beamproperties}) is a pulsed Nd:YAG, which can easily deliver up to 100 mJ of energy in a few ns.  Either the fundamental (1064 nm) or first harmonic (532 nm) modes are used, and neither seems to have a distinct advantage over the other, except for certain technical conveniences afforded by a visible laser.

Vapors of metals such as Na \cite{Maxwell2005} or Yb \cite{Patterson2007} can be easily produced by ablation of the solid metal; however, creation of gas phase molecules by laser ablation can be more complicated.  Molecules with a stable solid phase (such as PbO\cite{Maxwell2005}) can be created by simply using the solid phase as a precursor, but unstable molecules require careful choice of precursor.  It is often the case that the desired diatomic molecule MX has a stable solid form M$_a$X$_b$ which is a glass or ceramic, as is the case with BaF, CaH, SrF, YO, and ThO (and others).  In such cases, experience has shown that pressing (and sometimes sintering) a very fine powder of the stable solid often yields the best ablation target \cite{Hutzler2011,Barry2011}.

The most important advantage of laser ablation is that it can be used to create a very wide variety of atoms and molecules with high flux \cite{Campbell2009}.  The main drawback is that ablation is typically a ``violent,'' non-thermal process that can result in unusual behavior of the gases in the buffer cell, and in the resulting beam.  Ablation tends to give fairly inconsistent yields\cite{Hutzler2011,Barry2011}, and can create plasmas and complex plumes\cite{Gilgenbach1994} with temperatures of several thousand K\cite{Davis1985}.  However, buffer gas cells can be designed to allow proper thermalization (see Section \ref{sec:thermalization}) of the species, which can mitigate many of these problems.

\emph{Capillary filling.}  If the species of interest has an appreciable vapor pressure at convenient temperatures, one can simply flow the species into the cell through a gas fill line, just like the buffer gas\cite{Messer1984}.  This method is simple in principle, but in practice it introduces some technical challenges.  The cell and fill line must be carefully engineered to keep the fill line warm while keeping the cell cold and preventing freezing of the species at the point where the fill line enters the cell.  For species with an appreciable vapor pressure at room temperature, such as O$_2$ \cite{Patterson2007}, a properly-designed fill line can be attached directly to the buffer gas cell to flow the species.  For species that require higher temperatures to obtain an appreciable vapor pressure, for example a 600 K oven to produce atomic K vapor, a complex, multi-stage cell must be constructed \cite{Patterson2009} to keep heat from the oven out of the cell, as shown in Figure \ref{fig:capillary}.

\begin{figure*}[htbp]
	\centering
		\includegraphics[width=0.9\textwidth]{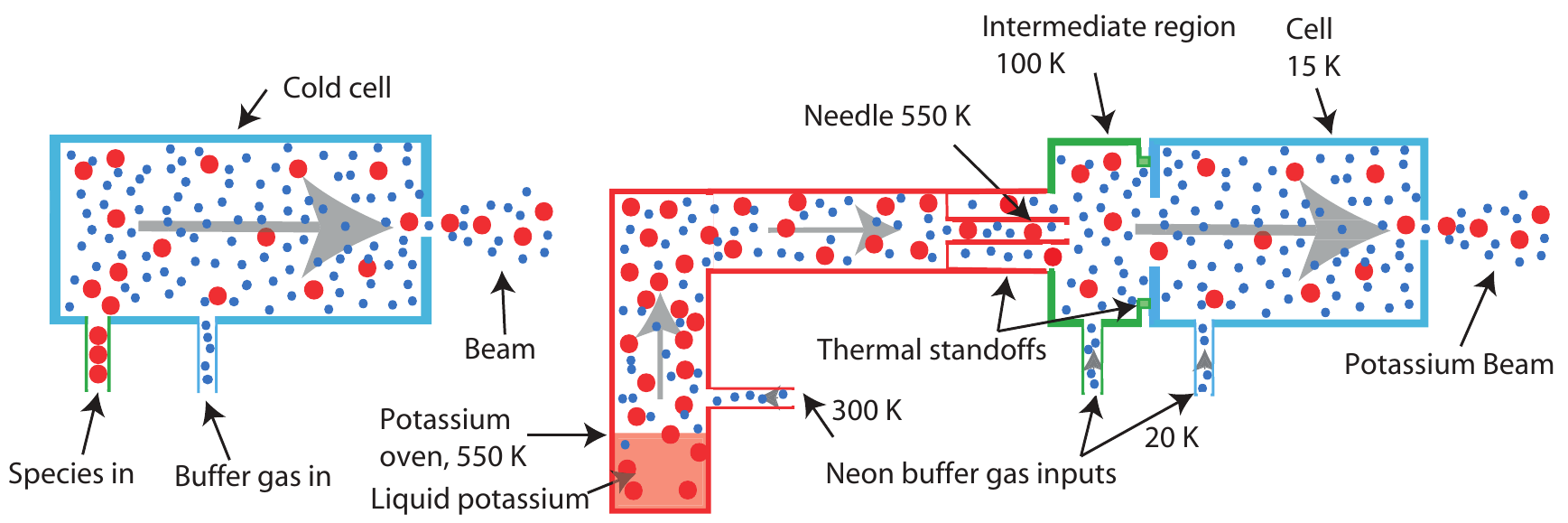}
	\caption{Left: simple capillary filling scheme for loading a molecule (large red circles) with vapor pressure at 300 K, such as O$_2$\cite{Patterson2007} or ND$_3$\cite{Patterson2009,Buuren2009}.  The species flows down a warm gas line and is mixed with the buffer gas (small blue circles), which flows down a cold gas line.  Right: Scheme for capillary injection of K atoms (large red circles) from a 550 K oven into a 15 K neon (small blue circles) buffer gas cell.  Unlike the simple capillary scheme, K atoms cannot simply flow down a warm gas line into the buffer gas cell, as the thermal conduction and blackbody heat loads would heat the cell too much.  Instead, a multi-stage cell with thermal standoffs heat links at each stage must be used to reduce the heat load on the cell to acceptable levels.  The amount of neon buffer gas introduced into the intermediate temperature regions must be sufficiently large to ensure that the K atoms are entrained in the flow and make it into the cold cell.  Reproduced from [Patterson, D., Rasmussen, J., \& Doyle, J. M., New J. Phys. 11(5), 055018 (2009) http://dx.doi.org/10.1088/1367-2630/11/5/055018] with permission from the Institute of Physics.}
	\label{fig:capillary}
\end{figure*}

The primary advantage of the capillary filling method is that it can be used to create high-flux beams that are continuous and robust.  The main drawback is the limited number of species of interest that have appreciable vapor pressure at temperatures low enough so that the method is feasible.  As the required temperature increases, the technical challenges rapidly multiply (see Figure \ref{fig:capillary}).

It should be noted that regardless of technique, the density of the species is typically (though not in the case of some capillary filling schemes\cite{Buuren2009,Sommer2009}) $<$ 1\% of the number density of the buffer gas.  This fact will be important later on, because it allows us to treat the the gas flow properties as being determined solely by the buffer gas, with the species as a trace component.

\subsubsection{Thermalization}\label{sec:thermalization}

Before the species flows out of the cell, it must undergo enough collisions with the cold buffer gas to become thermalized to the cell temperature.  A simple estimate of the necessary number of collisions can be obtained by approximating the species and buffer gases as hard spheres \cite{deCarvalho1999,Kim1997}.  The mean loss in kinetic energy of the species per collision with a buffer gas atom results in a mean temperature change of
\be \Delta T_s = -(T_s-T_b)/\kappa, \; \textrm{where} \; \kappa \equiv \frac{(m_b+m_s)^2}{2m_bm_s}. \ee
Here $T$ denotes temperature, $m$ denotes mass, and the subscripts ``$b$'' and ``$s$'' refer to the buffer gas and species of interest, respectively.  Therefore, the temperature of the species after $\mathcal{N}$ collisions, $T_s(\mathcal{N})$, will vary as
\be
T_s(\mathcal{N})-T_s(\mathcal{N}-1) = -(T_s(\mathcal{N}-1)-T_b)/\kappa
\ee
If we treat $\mathcal{N}$ as large and the temperature change per collision as small, we can approximate this discrete equation as a differential equation:
\be \frac{dT_s(\mathcal{N})}{d\mathcal{N}} = -(T_s(\mathcal{N})-T_b)/\kappa. \ee
Solving this differential equation yields the ratio between the species and buffer gas temperatures:
\bea
\frac{T_s(\mathcal{N})}{T_b} & = & 1+\left(\frac{T_s(0)}{T_b}-1\right)e^{-\mathcal{N}/\kappa} \\
& \approx & 1+\frac{T_s(0)}{T_b}e^{-\mathcal{N}/\kappa},
\eea
where in the last equality we have assumed that the species is introduced at a temperature much larger than the cell (and therefore buffer gas) temperature, i.e. $T_s(0) \gg T_b$.  If we estimate $m_b\sim 10$ amu, $m_s\sim 100$ amu, $T_b\sim 10$ K, and $T_s(0)\sim 1000$ K, the species should be within a few percent of the buffer gas temperature after $\sim$ 50 collisions.  Under typical cell geometry and flow conditions, the thermalization time is a few milliseconds, as shown in Figure \ref{fig:thermalization}.

\begin{figure}[htbp]
	\centering
		\includegraphics[width=0.45\textwidth]{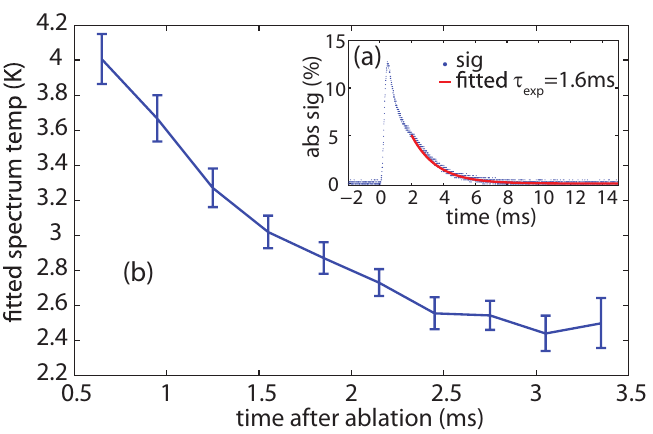}
	\caption{Thermalization of CaH in a helium buffer gas cell of temperature $\sim 2$ K\cite{Lu2011}.  (a) The absorption signal from a single ablation pulse.  (b) Temperature of the CaH molecules vs. time after ablation, determined by Doppler spectroscopy.  Reproduced from [Lu, H. et al., Phys. Chem. Chem. Phys. 13, 18986-18990 (2011) http://dx.doi.org/10.1039/c1cp21206k] by permission of the PCCP Owner Societies.}
	\label{fig:thermalization}
\end{figure}

While this simple model is good enough to obtain an estimate of the thermalization, it is not exact.  In analyzing thermalization of ablation-loaded YbF molecules in a helium buffer gas, Skoff et al. \cite{Skoff2011} found that the model above did not fit the data; however, replacing the constant buffer gas temperature with an initially larger temperature which then exponentially decays to the cell temperature resulted in good fits.  The physical motivation for this model is that the buffer gas is initially heated by the ablation, and then cools as the buffer gas re-thermalizes with the cell walls.  Skoff et al. present detailed analysis of thermalization in a buffer gas, and the reader is referred to their work for details.

Thus the cryogenic cell should be designed such that the species experiences at least 100 collisions or so before exiting the cell.  To find the thermalization length, i.e. the typical distance that the species will travel before being thermalized, we need the mean free path of the species in the buffer gas, which is given approximately by\cite{Hasted1972}
\be
\lambda_{s-b,0} = \frac{(n_{0,b}\sigma_{b-s})^{-1}}{\sqrt{1+m_s/m_b}}
\approx \frac{ A_{aperture}\bar{v}_{0,b}}{4 f_{0,b} \sigma_{b-s} \sqrt{m_s/m_b}}
\label{mfp}\ee
where $\sigma_{b-s}$ is the thermally averaged elastic collision cross section (since the cross section typically varies with temperature), we assume $m_s \gg m_b$, and have used Eq. (\ref{flowtodensity}) in the second step.  For typical values of $\sigma_{b-s}\approx 10^{-14}$ cm$^2$, this mean free path is $\sim 0.1$ mm.  Therefore, the thermalization length for the species in the buffer gas cell is typically no more than $100\times 0.1$ mm = 1 cm.

Note that the above discussion has pertained only to translational temperatures, yet buffer gas cooling is also effective at thermalization of internal states.  Typical rotational relaxation cross sections for molecules with helium buffer gas are typically of order $\sigma_{rot}\sim 10^{-(15-16)}$ cm$^{2}$, which means that around $\sigma_{b-s}/\sigma_{rot}\sim 10-100$ collisions are required to relax (or ``quench") a rotational state\cite{Campbell2009}.  Since this is a typical number of collisions required for motional thermalization, buffer gas cooling can be used to make samples of molecules which are both translationally and rotationally cold.  Vibrational relaxation is less efficient, since the cross sections for vibrational quenching are typically several order of magnitude smaller than those for translational or rotational relaxation\cite{Campbell2009}.  Several experiments\cite{Weinstein1998,Campbell2008,Barry2011} have seen the vibrational degree of freedom not in thermal equilibrium with the rotational or translational degrees of freedom.  Further discussion of internal relaxation of molecules may be found elsewhere\cite{Campbell2009,Scoles1988,Pauly2000}.

\subsubsection{Diffusion}

Once the species of interest is introduced into the cell and thermalized, we must consider the diffusion of the species in the buffer gas.  Understanding the diffusion is crucial since the buffer gas cell is typically kept at a temperature where the species of interest has essentially no vapor pressure, and if it is allowed to diffuse to the cell walls before exiting the cell it will freeze and be lost.  The diffusion constant for the species diffusing into the buffer gas is\cite{Hasted1972}
\be D = \frac{3}{16(n_{0,s}+n_{0,b})\sigma_{b-s}}\left(\frac{2\pi k_B T_0}{\mu}\right)^{1/2}, \ee
where $\mu = m_sm_b/(m_s+m_b)$ is the reduced mass.  It should be noted that different works use different forms for the diffusion coefficient \cite{Campbell2009}, though they agree to within factors of order unity and are all suitable for making estimates.  Making the approximations $n_{0,s}\ll n_{0,b}$ and $m_s\gg m_b$, we find
\be D = \frac{3}{16n_{0,b}\sigma_{b-s}}\left(\frac{2\pi k_B T_0}{m_b}\right)^{1/2} = \frac{3\pi}{32}\frac{\bar{v}_{0,b}}{n_{0,b}\sigma_{b-s}}, \ee
After a time $t$, a species molecule will have a mean-squared displacement of
\be   \langle\Delta x^2\rangle(t) = 6 D t = \frac{9\pi}{16}\frac{\bar{v}_{0,b}}{n_{0,b}\sigma_{b-s}}\;t \ee
from its starting point.\cite{Pathria1996}  Since the characteristic length of the cell interior is the cross-sectional length $d_{cell}$, we can define the diffusion timescale $\tau_{dif\! f}$ as $\langle\Delta x^2\rangle(\tau_{dif\! f}) = d_{cell}^2\approx A_{cell}$, or
\be \tau_{dif\! f} = \frac{16}{9\pi} \frac{A_{cell}n_{0,b}\sigma_{b-s}}{\bar{v}_{0,b}}. \label{taudiff}\ee
The diffusion time is typically 1-10 ms.

Skoff et al. \cite{Skoff2011} performed a detailed theoretical analysis and compared the results to measured absorption images in order to understand diffusion of YbF and Li in a helium buffer gas.  The diffusion behavior was studied as a function of both helium density and cell temperature for both species.  At low densities they verified the linear relationship between $\tau_{dif\!f}$ and $n_{0,b}$ from Eq. (\ref{taudiff}), and a departure from linearity at high density.  While this was not the first time that this relationship was observed\cite{Sushkov2008}, their data and analysis allows for an informative possible explanation: At low buffer gas densities, the ablation ejects the YbF molecules so that they are distributed all across the cell, and the diffusion model discussed above is valid.  At high density, the YbF molecules are localized near the target and therefore diffuse only through higher-order modes with shorter timescales.  Previous models for this behavior included formation of dimers or clusters\cite{Sushkov2008}; however, the model proposed by Skoff et al.\cite{Skoff2011} seems reasonable considering that this behavior has been seen for several species, including both atoms and molecules\cite{Weinstein1998,Sushkov2008,Lu2008,Lu2009II,Skoff2011}

\subsubsection{Extraction from the buffer cell}\label{sec:extraction}

So far we have not considered the beam at all, having focused entirely on the in-cell dynamics.  Once the species of interest is created in the gas phase and cooled in the buffer gas cell, it is necessary that the species flow out of the cell so that it can create a beam.  As discussed above, extraction of the species from the cell must occur faster than the diffusion timescale $\tau_{dif\! f}$, so let's work out an estimate of the extraction, or ``pumpout'' time.  The rate at which the buffer gas out of the cell is given by the molecular conductance of the cell aperture,

\be   \dot{N}_b = \frac{1}{4} N_{b}\bar{v}_{0,b}A_{aperture}/V_{cell},\ee
where $N_{b}$ indicates the total number of buffer gas atoms in the cell, and $\dot{N}_b$ is the rate at which they are flowing out of the cell.  The solution is an exponential decay with timescale $\tau_{pump}$, the pumpout time, given by
\be   \tau_{pump} = \frac{4 V_{cell}}{\bar{v}_{0,b}A_{aperture}}. \ee

The pumpout time is typically around 1-10 ms.  Note that the pumpout time also sets the duration of the molecular pulse in the case of a beam with pulsed loading.  If the buffer gas density in the cell is high enough that the species of interest follows the buffer gas flow, then this is a good estimate for the pumpout time for the species of interest as well.  We now define\cite{Patterson2007} a dimensionless parameter to characterize the extraction behavior of the cell
\be   \gamma_{cell} \equiv \frac{\tau_{dif\! f}}{\tau_{pump}} = \frac{4}{9\pi} \frac{n_{0,b}\sigma_{b-s}A_{aperture}}{L_{cell}} \approx \frac{\sigma_{b-s}f_{0,b}}{L_{cell}\bar{v}_{0,b}}, \label{gammacell}\ee
where in the last approximation we dropped the order-unity prefactor as this is simply an estimate.  This parameter $\gamma_{cell}$ characterizes the extraction behavior, and can be divided into two limits (see Figure \ref{fig:extraction}).

For $\gamma_{cell}\lesssim 1$, the diffusion to the walls is faster than the extraction from the cell, so the majority of species molecules will stick to the cell walls and be lost.  This ``diffusion limit'' \cite{Patterson2007} is characterized by low output flux of the molecular species, and is typically accompanied by a velocity distribution in the beam that is similar to that inside the cell \cite{Maxwell2005}.  In this limit, increasing the flow (thereby increasing $\gamma_{cell}$) has the effect of increasing the extraction efficiency $f_{ext}$, defined as the fraction of molecules created in the cell which escape into the beam.  The precise dependence of $f_{ext}$ on $\gamma_{cell}$  is highly variable, and has been observed to be approximately linear, exponential, or cubic for the various species which have been examined\cite{Maxwell2005,Patterson2007,Hutzler2011,Barry2011}.

For $\gamma_{cell}\gtrsim1$, the molecules are mostly extracted from the cell before sticking to the walls, resulting in a beam of increased brightness.  This limit, called ``hydrodynamic entrainment'' or ``hydrodynamic enhancement'' \cite{Patterson2007}, is characterized by high output flux of the molecular species, and is typically accompanied by velocity distribution which can vary considerably from that present inside the cell.  In this regime, the extraction efficiency plateaus and can be as high as $>$40\%\cite{Patterson2007,Barry2011}, but is typically observed to be around 10\%.

\begin{figure}[htbp]
	\centering
		\includegraphics[width=.45\textwidth]{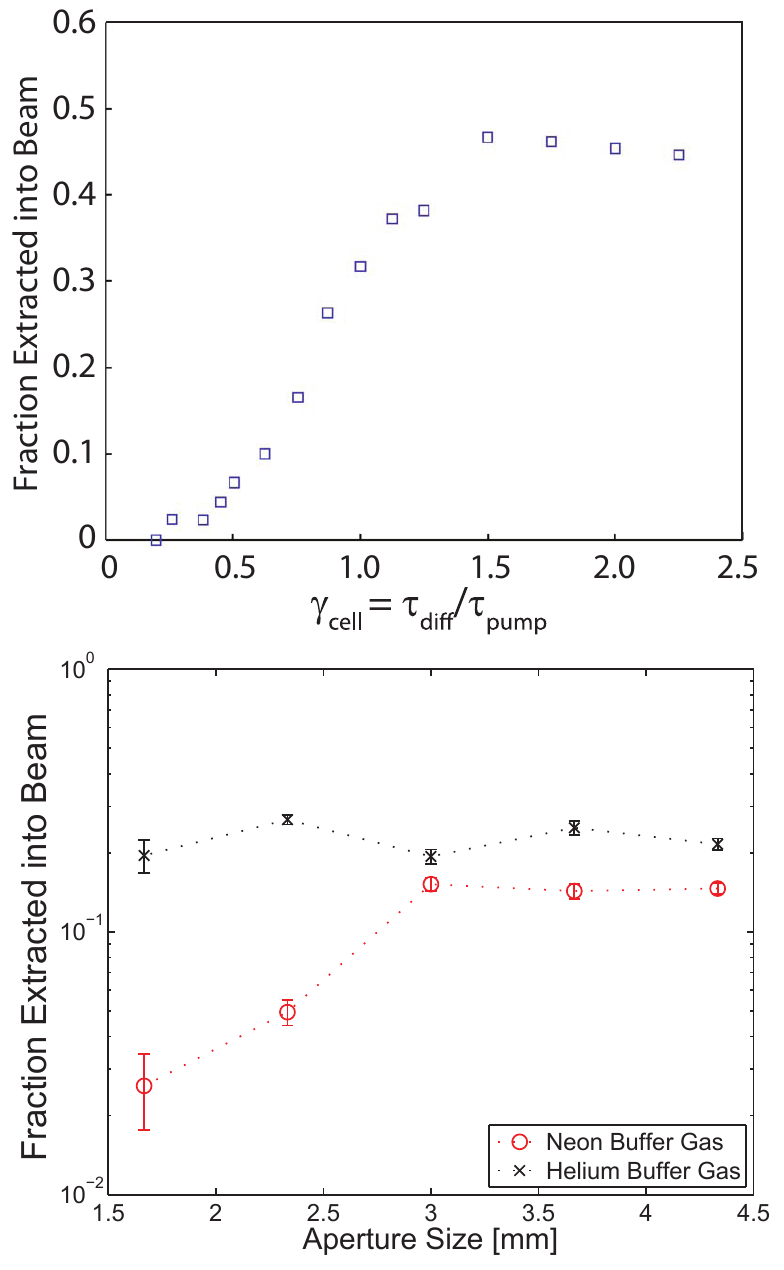}
	\caption{Extraction from a buffer gas cell.  Top: Fraction of Yb atoms extracted into a helium cooled buffer gas beam vs. the parameter $\gamma_{cell}$ (Eq. (\ref{gammacell})).  Reproduced from [Patterson, D. \& Doyle, J. M., J. Chem. Phys. 126, 154307 (2007) http://dx.doi.org/10.1063/1.2717178].  Copyright (2007), American Institute of Physics.  Bottom: Fraction of ThO molecules extracted into a helium ($\times$) and neon ($\circ$) buffer gas vs. the cell aperture size with a fixed flow rate.  Reproduced from [Hutzler, N. et al., Phys. Chem. Chem. Phys. 13, 18976-18985 (2011) http://dx.doi.org/10.1039/c1cp20901a (2011)] by permission of the PCCP Owner Societies.}
	\label{fig:extraction}
\end{figure}

It should be noted that while the parameter $\gamma_{cell}$ can be used to estimate the extraction efficiency quite well \cite{Patterson2007,Hutzler2011,Barry2011}, there have been instances where this simple estimate breaks down.  According to Eq. (\ref{gammacell}), the parameter $\gamma_{cell}$ has no explicit dependence on the aperture diameter; however, Hutzler et al. \cite{Hutzler2011} found that while varying the cell aperture diameter \emph{in situ} without varying other parameters, the extraction efficiency of a ThO beam in a neon buffer gas began to decrease with decreasing aperture size, as shown in Figure \ref{fig:extraction}.  These measurements suggest that the cell aperture diameter should not be too small ($\lesssim$ 3 mm) in order to achieve good extraction.

Regardless of the value for $\gamma_{cell}$, thermalization (see Section \ref{sec:thermalization}) must occur on a timescale faster than either $\tau_{pump}$ or $\tau_{dif\!f}$ to cool the species.  Since neither $\tau_{dif\!f}$ nor $\tau_{pump}$ impose strict constraints on the cell geometry, it is possible to have both good thermalization and efficient extraction, as demonstrated by the large number of high flux, cold beams created with the buffer gas method (see Table \ref{table:beamproperties}).

\subsection{Properties of buffer gas cooled beams}

\begin{table*}
\centering
\begin{tabular}{llll}
Species 	& Output/Brightness					& $v_{||}$[m s$^{-1}$] & $\Delta v_\|[$m s$^{-1}$] \\
\hline \emph{Molecules} \\
BaF\cite{Rahmlow2010}			  												& $1.6\times10^{11}$ sr$^{-1}$ pulse$^{-1}$				&  --										& --	\\
CaH\cite{Lu2011}$^\textrm{A}$			  							& $5-500\times 10^{8}$ sr$^{-1}$ pulse$^{-1}$	& 40--95							& 65 	\\
CH$_3$F\cite{Sommer2009}$^\textrm{B}$	  				& -- 												& 45									& 35 	\\
CF$_3$H	\cite{Sommer2009}$^\textrm{B}$	  				& -- 												& 40									& 35 	\\
H$_2$CO\cite{Buuren2009}$^\textrm{B}$						& --												& --									& --	\\
ND$_3$\cite{Patterson2009}$^\textrm{B}$  				& $3-200\times10^{8}$ s$^{-1}$			&	60--150							& 25--100 	\\
ND$_3$\cite{Buuren2009,Sommer2009}$^\textrm{B}$ 	& $1-10\times 10^{10}$ s$^{-1}$		& 65									& 50 	\\
ND$_3$\cite{Sawyer2011}$^\textrm{B}$						  & $1\times 10^{11}$ s$^{-1}$				& 100									& -- 	\\
O$_2$\cite{Patterson2007}$^\textrm{B}$				  	& $3\times 10^{12}$ s$^{-1}$				& --									& -- 	\\
PbO\cite{Maxwell2005}												  			& $3-100\times 10^8$ sr$^{-1}$ pulse$^{-1}$		& 40--80							& 30--40 	\\
SrF\cite{Barry2011}						  										& $1-12\times10^{10}$ sr$^{-1}$ pulse$^{-1}$	& 125--200						& 60--80 	\\
SrO\cite{Petricka2007}			  											& $3-100\times 10^9$ sr$^{-1}$ pulse$^{-1}$		& 65--180							& 35--50 	\\
ThO\cite{Hutzler2011}					  										& $1-100\times10^{10}$ pulse$^{-1}$		& 120--200						& 30--45 	\\
YbF\cite{Skoff2011,Hendricks2011PC}									& --												&  --									& -- 	\\
YO\cite{Hummon2011PC}				 												&	--												&	160									&	-- 	\\
\emph{Atoms} \\
K\cite{Patterson2009}$^\textrm{D}$				  				& $1\times 10^{16}$ sr$^{-1}$ s$^{-1}$		&  130	 							& 120 	\\
Na\cite{Maxwell2005}					  										& $2-400\times 10^8$ sr$^{-1}$ pulse$^{-1}$		&  80--135						& 60-120 	\\
Rb\cite{Lu2009}$^\textrm{C,D}$								  		& $3\times 10^{10}$ s$^{-1}$				& 190									& 25--30 	\\
Yb\cite{Patterson2007}				  										& $5\times 10^{13}$ sr$^{-1}$ pulse$^{-1}$		& 130									& -- 	\\
Yb\cite{Patterson2007}$^\textrm{A}$					  		& $5\times 10^{10}$ pulse$^{-1}$				& 35									& -- 	\\
\end{tabular}
\caption{A list of molecules which have been cooled in buffer gas beams, along with any measured properties.  The output is either reported as molecules per state per pulse for pulsed beams, or per second for continous beams.  If an angular spread was measured, the brightness is reported as molecules per state per unit steradian (sr) per pulse or per second.  $v_{\|}$ is the forward velocity and $\Delta v_{\|}$ is the full-width at half-maximum of the forward velocity distribution.  Beams are ablation loaded and pulsed unless otherwise noted.  A dash (--) means that the beam property was not reported in the indicated references. Notes: A = slowing cell (see Section \ref{sec:slowingcells}), B = continuous beam, velocity selected or guided by electromagnetic fields, C = flux reported is after beam collimation, D = continuous beam, loaded via capillary.}
\label{table:beamproperties}
\end{table*}

Now that we have reviewed the requirements for introduction of the species, thermalization, and extraction, we may focus on the properties of the resulting beam.

\subsubsection{Characterization of gas flow regimes}

As mentioned earlier, the species of interest typically constitutes $<1\%$ of the number density of the buffer gas, so the flow properties are determined entirely by the buffer gas.  In this section we will introduce the Reynolds number, which we will use to characterize gas flow.  Note that the Reynolds (and Knudsen) number refers exclusively to the buffer gas, and not the species of interest (for our treatment).  For a more thorough discussion about gas flow, see texts about beams and gas dynamics\cite{Scoles1988,Pauly2000,Sone2007}.

The Reynolds number is defined as the ratio of inertial to viscous forces in a fluid flow\cite{Fox1998,VonKarman1963}
\be   \re = \frac{F_{intertial}}{F_{viscous}} = \frac{\rho w^2d^2}{\mu w d} = \frac{\rho w d}{\mu},\ee
where $\rho$ is the density, $w$ is the flow velocity, $\mu$ is the (dynamic) viscosity, and $d$ is a characteristic length scale.  In terms of kinetic quantities, we make use of the relationship between mean free path, density, and viscosity\cite{Sone2007,VonKarman1963}
\be   \mu \approx \frac{1}{2}\rho\lambda\bar{v}, \ee
where $\bar{v}$ is the mean thermal velocity and $\lambda$ is the mean free path, to express the Reynolds number as
\be   \re = \frac{\rho w}{\frac{1}{2}\rho\lambda\bar{v}/d} \approx 2\frac{w/\bar{v}}{\lambda/d} \approx 2\frac{\ma}{\kn},  \ee
where $w/\bar{v}\approx\ma$ is the Mach number, and $\kn = \lambda/d$ is the Knudsen number.  The Mach number for a gas of atomic weight $m$ is defined as $\ma=w/c$, where
\be   c\equiv \sqrt{\frac{\gamma k_B T}{m}} \label{speedofsound} \ee
is the speed of sound in the gas, and $\gamma$ is the specific heat ratio.  The value of $\gamma$ for a monoatomic gas (the relevant case for buffer gas beams) is $\gamma=5/3$, so in this case
\be c = \bar{v}\sqrt{\frac{5\pi}{8}} \approx 0.8 \bar{v}, \ee
and we are justified in approximating $\ma\approx w/\bar{v}$.  This gives us the important von K\'{a}rm\'{a}n relation \cite{Sone2007,VonKarman1963}
\be   \ma \approx \frac{1}{2}\kn\re. \ee

To relate these quantities to the case under consideration, we must mention two important facts about gas flow from an aperture into a vacuum \cite{Scoles1988}.  First, the most relevant geometrical length scale which governs the flow behavior is the aperture diameter, since the properties of the beam are set almost entirely by collisions that occur near the aperture.  Therefore the characteristic length scale appearing in the formula for $\re$ should be $d=d_{aperture}$.  Second, near the aperture the gas atoms are traveling near their mean thermal velocity, so $\ma\approx 1.$  Combining these facts with Eqs. (\ref{flowtodensity}) and (\ref{mfp}) allows us to write the relevant expression for $\re$ in our situation,
\be
 \re  \approx  \frac{2d_{aperture}}{\lambda_{b-b,0}} \approx \frac{8\sqrt{2} f_{0,b} \sigma_{b-b}}{ d_{aperture}\bar{v}_{0,b}}.
\label{knre} \ee

We shall parameterize the beam by the Reynolds number $\re$ of the buffer gas flow at the aperture.  According to Eq. (\ref{knre}), $\re$ is approximately $2\kn^{-1}=2d_{aperture}/\lambda_{b-b}$, or twice the number of collisions within one aperture diameter of the aperture, i.e. ``near'' the aperture.  From Eqs. (\ref{knre}) and (\ref{flowtodensity}) we can see that the in-cell buffer gas density, buffer gas flow rate, Reynolds number, and number of collisions near the aperture are all linearly related.  The possible types of flow can be roughly divided into three Reynolds number regimes, each of which will be discussed in the remainder of this section:
\begin{itemize}
\item \emph{Effusive regime, $\re\lesssim 1$}:  In this regime there are typically no collisions near the aperture, so the beam properties are simply a sampling of the thermal distribution present in the cell.  This regime is discussed further in Section \ref{sec:effusive}.
\item \emph{Intermediate, or partially hydrodynamic regime, $1\lesssim \re\lesssim 100$}: Here there are enough collisions near the aperture to change the beam properties from those present in the cell, but not enough so that the flow is fluid-like.  Buffer gas beams typically operate in this regime; however, we will see examples of buffer gas beams in all three regimes.
\item \emph{Fully hydrodynamic, or ``supersonic'' regime, $100 \lesssim \re$}: In this regime the buffer gas begins to behave more like a fluid, and the beam properties become similar to those of a beam cooled via supersonic expansion.  This regime is discussed further in Section \ref{sec:supersonic}.
\end{itemize}

\subsubsection{Forward velocity}\label{sec:forwardvelocity}

\begin{figure}[htbp]
	\centering
		\includegraphics[width=0.45\textwidth]{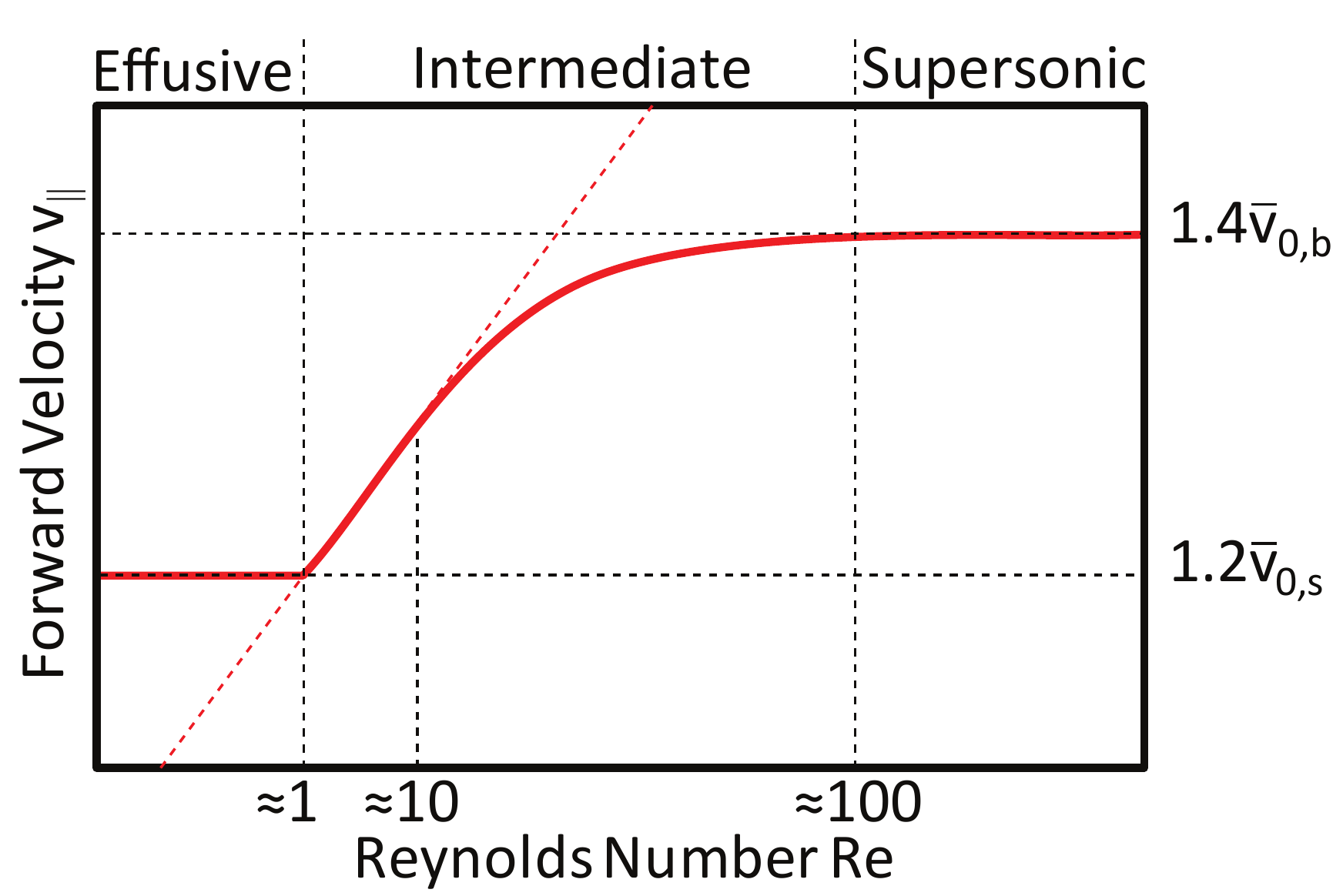}
	\caption{A schematic representation of beam forward velocity vs. Reynolds number.  In the effusive regime $(\re\lesssim 1)$, the forward velocity is the thermal velocity of the (heavy) species.  In the intermediate regime, collisions of the species with the buffer gas near the aperture accelerate the species; the velocity increase is linear with the Reynolds number until $\re\approx 10$, and then begins to asymptote to the final value.  In the supersonic regime $(\re\gtrsim 100)$, the species has been fully accelerated to the forward velocity of the (light) buffer gas.  Each of these regimes is discussed in detail below.}
	\label{fig:vf_vs_re_schematic}
\end{figure}

If the beam is in the effusive regime, then there are typically no collisions near the aperture, and the forward velocity of the beam $v_{\|,s}$ is given by the forward velocity of an effusive beam (Eq. (\ref{effvelocity})), where the appropriate thermal velocity is that of the species, i.e.
\be   v_{\|,s} \approx \frac{3\pi}{8}\bar{v}_{0,s} \approx 1.2 \bar{v}_{0,s} \qquad (\re\lesssim 1).\ee
As we will discuss, creating a purely effusive buffer gas beam without using advanced cell geometries (Section \ref{sec:slowingcells}) can be challenging.

In the intermediate regime, the molecules undergo collisions with the buffer gas atoms near (i.e. within one aperture diameter of) the cell aperture, whose average velocity is $\bar{v}_{0,b}$.  This is larger than that of the (typically) heavier species $\bar{v}_{0,s}$ by a factor of $\sqrt{m_s/m_b}$.  Since the collisions near the aperture are primarily in the forward direction, the species can be accelerated, or ``boosted,'' to a forward velocity, $v_{\|,s}$, which is larger than the thermal velocity of the molecules (just as with supersonic beams\cite{Scoles1988}).

If there are few collisions, we can estimate the relationship between $\re$ and $v_{\|,s}$ with a simple model from Maxwell et al.\cite{Maxwell2005}.  Near the aperture, the molecules undergo approximately $\re/2$ collisions.  Each of these collisions gives the molecules a momentum kick in the forward direction of about $\approx m_{b} v_{b}$, so the net velocity boost is $\approx v_{b} m_{b}\re/2m_{mol}$.  Since there are a small number of collisions, the forward velocity of the buffer gas is approximately given by the mean forward velocity of an effusive beam (Eq. (\ref{effvelocity})), $v_{\|,b}\approx 1.2 \bar{v}_{0,b}$, so for $1\lesssim \re \lesssim 10$ (we shall justify the $\re\lesssim 10$ cutoff later on),
\be  v_{\|,s} \approx 1.2\bar{v}_{0,s}+ 0.6\bar{v}_{0,b}\re \frac{m_{b}}{m_{mol}} . \ee
Therefore, the forward velocity should increase linearly with $\re$ (and therefore with in-cell buffer gas density, or buffer gas flow).  Several papers \cite{Maxwell2005,Hutzler2011,Barry2011} have considered the behavior of $v_{\|,s}$ vs. $\re$, and have seen this linear dependence at low $\re$.  The more recent papers \cite{Hutzler2011,Barry2011} show data where the Reynolds number is large enough that a departure from the linear regime can be seen.  Hutzler et al. \cite{Hutzler2011} calculated the slope of the $v_{\|,s}$ vs. $\re$ relationship at low flow and found good agreement with the above model.

This model necessarily breaks down as $v_{\|,s}$ approaches $\sim\bar{v}_{0,b}$, since the maximum possible forward velocity is $1.4\bar{v}_{0,b}$ (from Eq. (\ref{finalvf})), the forward velocity of a fully hydrodynamic buffer gas expansion.  We therefore expect that the forward velocity should saturate to this value at large enough $\re$.  To model the behavior outside of the linear regime considered above, Hutzler et al. \cite{Hutzler2011} used the ``sudden freeze'' model \cite{Pauly2000}, where it is assumed that the species molecules are in equilibrium with the buffer gas until the point along the beam where the buffer density is decreased enough such that there are no more collisions and the beam properties stop changing, or ``freeze.''  The functional form for this model is (for $\re\gtrsim 10$)
\be v_{\|,s}(\re) \approx   1.4\bar{v}_{0,b}\sqrt{1-4\re^{-4/5}}, \label{sfvelocity} \ee
and fits the data fairly well\cite{Hutzler2011}.  The estimate for where the linear to sudden-freeze model transition occurs is by finding at which flow there are collisions at a distance larger than one aperture diameter from the aperture, which happens for $\re\gtrsim 10$ \cite{Hutzler2011}.

Finally, for large enough $\re$, there should be enough collisions to fully boost the molecules to the forward velocity of the buffer gas,
\be   v_{\|,s} \approx v_{\|,b} \approx 1.4 \bar{v}_{0,b} \qquad (\re\gtrsim 100) \ee
where the cutoff $\re\gtrsim 100$ means that  $v_{\|,s}(100) \approx 95\%\times v_{\|,b}$ according to Eq. (\ref{sfvelocity}), and from experiment \cite{Hutzler2011,Barry2011}.  This limit corresponds to the supersonic flow regime.

\begin{figure}[htbp]
	\centering
		\includegraphics[width=0.45\textwidth]{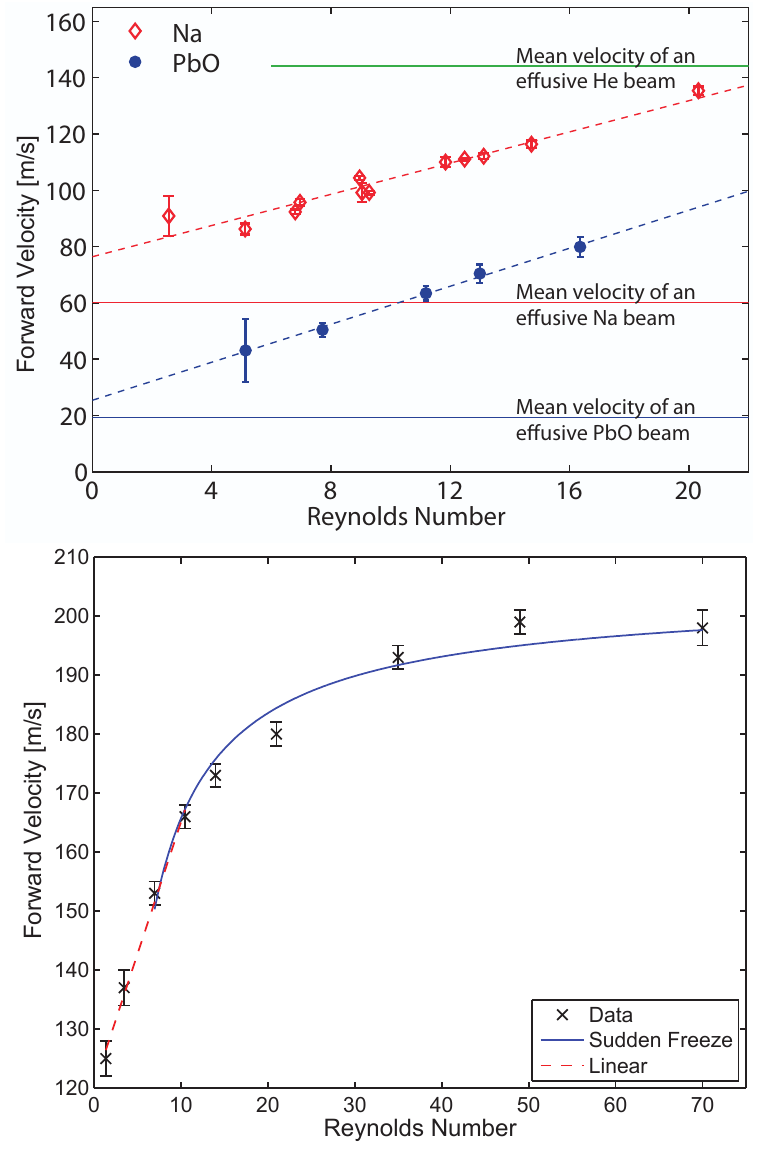}
	\caption{Forward velocity of extracted beam vs. buffer gas flow rate.  Top: Forward velocity for Na and PbO in a helium cooled buffer gas beam. Reproduced from [Maxwell et al., Phys. Rev. Lett. Vol. 95, 173201 (2005) http://dx.doi.org/10.1103/PhysRevLett.95.173201]  Copyright (2005) by the American Physical Society.  Bottom: Forward velocity for ThO in a neon cooled buffer gas beam.  Here the beam starts in the linear regime, and then fully saturates at the neon supersonic forward velocity.  Reproduced from [Hutzler, N. et al., Phys. Chem. Chem. Phys. 13, 18976-18985 (2011) http://dx.doi.org/10.1039/c1cp20901a (2011)] by permission of the PCCP Owner Societies.}
	\label{fig:vf_vs_flow}
\end{figure}

\subsubsection{Velocity spreads}

\emph{Forward, or longitudinal velocity spread.}  In the effusive regime, the longitudinal velocity spread is the spread (FWHM) of the thermal 1D Maxwell-Boltzmann distribution,
\be   \Delta v_{\|,s} = \sqrt{\frac{8\ln(2)k_bT_0}{m_s}} = \bar{v}_{0,s}\sqrt{\pi\ln(2)} \approx 1.5\bar{v}_{0,s}.  \label{effforward}\ee
Maxwell et al.\cite{Maxwell2005} operated a buffer gas beam of PbO in this regime and found that the longitudinal velocity spread (as well as the transverse 
velocity spread, and rotational level distribution) were all in agreement with a thermal distribution at the cell temperature.

If the Reynolds number of the flow is high enough, then the forward velocity spread can begin to decrease due to isentropic expansion of the buffer gas into the vacuum region.  The translational temperature in the longitudinal (i.e. forward) direction can be decreased below the cell temperature, as will be discussed in Section \ref{sec:isentropicexpansion}.

\emph{Transverse velocity spread}.  Similarly, the transverse spread in the effusive regime is
\be   \Delta v_{\perp,s} \approx 1.5\bar{v}_{0,s}.  \label{efftransverse}\ee

Also similar to the case with forward velocity, in the intermediate regime there will be collisions between the species and buffer gas near the aperture which can increase the transverse velocity spread.  For low $\re$, we can estimate the transverse velocity spread near the aperture with a that is similar to the model for forward velocity,
\be  \Delta v_{\perp,s} \approx 1.5\bar{v}_{0,s} + (\re/2) \bar{v}_{0,b}\frac{d_{cell}^2}{d_{aperture}^2} \frac{m_{b}}{m_{mol}}.\ee
Hutzler et al. \cite{Hutzler2011} used this model to describe the behavior of the transverse spread of a ThO beam in neon buffer gas and found it approximated the ratio $\Delta v_{\perp,s}/\re$ well.

Experimentally, we are less concerned with the transverse velocity spread near the aperture than with the final velocity spread downstream, after collisions have ceased.  This is more complicated to model since it conflates the dynamics discussed above with the dynamics of the expansion.  However, a similar argument shows that the transverse spread should increase linearly (above some $\re$) with increasing $\re$.  This was examined in detail\cite{Hutzler2011,Barry2011}, and a linear relationship was seen, with a ratio $\Delta v_{\perp,s}/\re$ that was $\sim 2$ times the ratio near the aperture\cite{Hutzler2011}.

Finally, as $\re$ is further increased, the transverse velocity spread should saturate at the transverse spread of the buffer gas.  Barry et al.\cite{Barry2011} started to see this saturation behavior for SrF in a helium buffer gas start around $\re\approx 100$.

We have seen that the shape of the $\re$ vs. $\Delta v_{\perp,s}$ relationship (see Figure \ref{fig:vt_vs_re}) is very similar to that which is plotted in Figure \ref{fig:vf_vs_re_schematic}.  However, as we shall see in the following section, the Reynold's numbers where the transitions occur do not have to be the same as in the case of forward velocity.

\begin{figure}[htbp]
	\centering
		\includegraphics[width=0.45\textwidth]{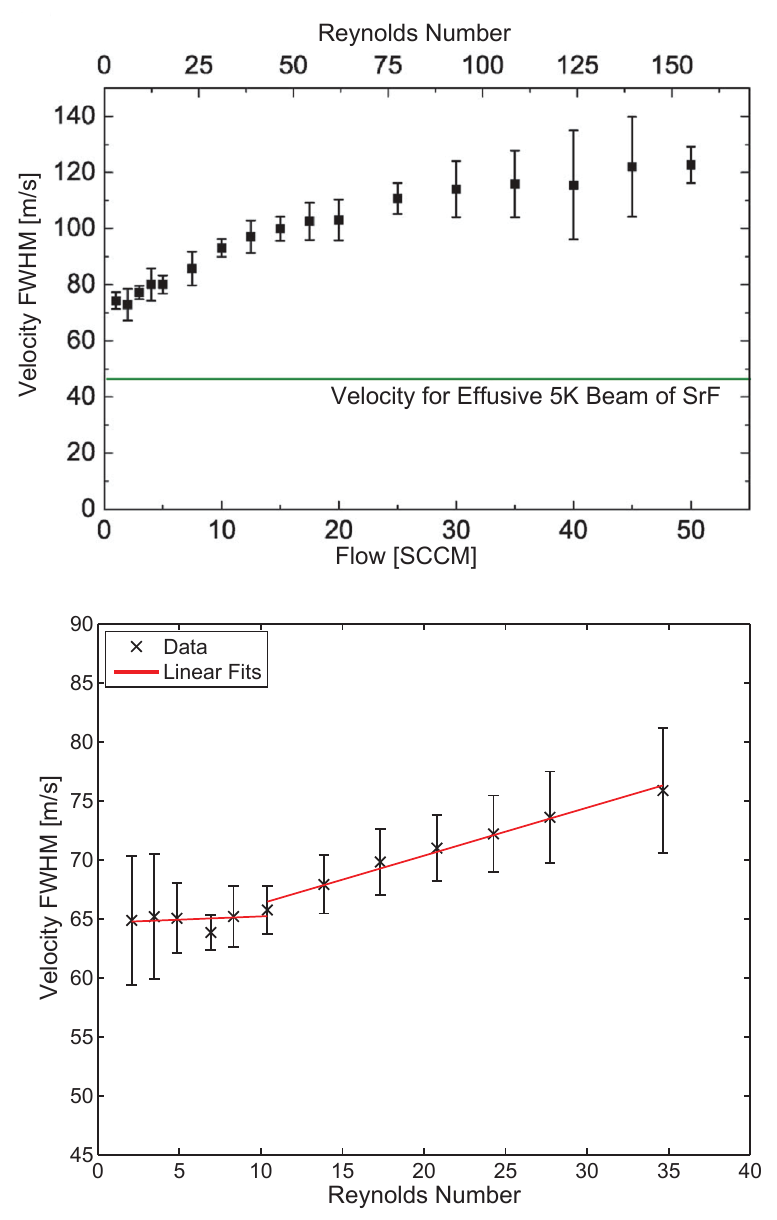}
	\caption{Behavior of transverse velocity spread vs. Reynolds number.  Top: transverse velocity spread of SrF in a helium buffer gas cooled beam vs. $\re$ at a fixed distance of 20 mm ($\approx 7 d_{aperture}$) from the cell aperture\cite{Barry2011}.  At low $\re$ the transverse spread is in the linear regime, but it begins to saturate around $\re\approx 50$.  Reproduced from [Barry et al., Phys. Chem. Chem. Phys. 13, 18936-18947 (2011) http://dx.doi.org/10.1039/c1cp20335e] by permission of the PCCP Owner Societies.  Bottom: transverse velocity spread of ThO in a neon buffer gas\cite{Hutzler2011} vs. $\re$ at a fixed distance of 1 mm ($<d_{aperture})$ from the cell aperture.  The transition from flat (and approximately equal to the spread from a thermal distribution at $T_0$) to linearly increasing occurs around $\re\approx 10$.  The data is well described by the model mentioned in the text.  Reproduced from [Hutzler, N. et al., Phys. Chem. Chem. Phys. 13, 18976-18985 (2011) http://dx.doi.org/10.1039/c1cp20901a] by permission of the PCCP Owner Societies.}
	\label{fig:vt_vs_re}
\end{figure}

Typically, the transverse velocity spread is not a directly important parameter.  Actual experiments performed with atomic and molecular beams are generally performed at a distance from the cell aperture that is many times larger than $d_{aperture}$, and only sample a very small solid angle of the output beam.  This is both to reduce the gas load on the vacuum apparatus (the unused portions of the beam can be differentially pumped to maintain a good vacuum), and to select atoms or molecules with nearly-identical directions of flight to reduce Doppler broadening of transversely probed spectroscopic transitions.  However, the transverse velocity spread is important because it factors in to the divergence of the molecular beam, which will be discussed in the next section.

\subsubsection{Angular spread and divergence}\label{sec:divergence}

As mentioned earlier, the experimentally useful portion of an atomic or molecular beam is the portion that makes it through a typically small detection region at a large distance away.  For this reason, an important characteristic of a beam is the value of the angular density distribution of the molecular beam.  A parameter that can be used to characterize the angular spread of the beam is the full-width at half-maximum (FWHM) of the angular distribution, $\Delta\theta$.  This parameter is defined by equating ${\dot{N}(\Delta\theta)}=\frac{1}{2}{\dot{N}_{total}}$, where $N(\theta)$ is the density at an angle of $\theta$ from the aperture normal at a fixed distance from the aperture (see Section \ref{sec:effusive}).  This parameter is very simple to calculate from transverse and longitudinal Doppler spectroscopic data; if a beam has a transverse velocity spread $\Delta v_{\perp}$ and forward velocity $v_\|$, then (see Figure \ref{fig:angles_schematic})
\bea
\tan(\Delta\theta/2) & = & \frac{\Delta v_\perp/2}{v_\|} \\
\Rightarrow \Delta\theta & = & 2\arctan\left(\frac{\Delta v_\perp/2}{v_\|}\right)
\label{thetafwhm}\eea

\begin{figure}[htbp]
	\centering
		\includegraphics[width=0.5\textwidth]{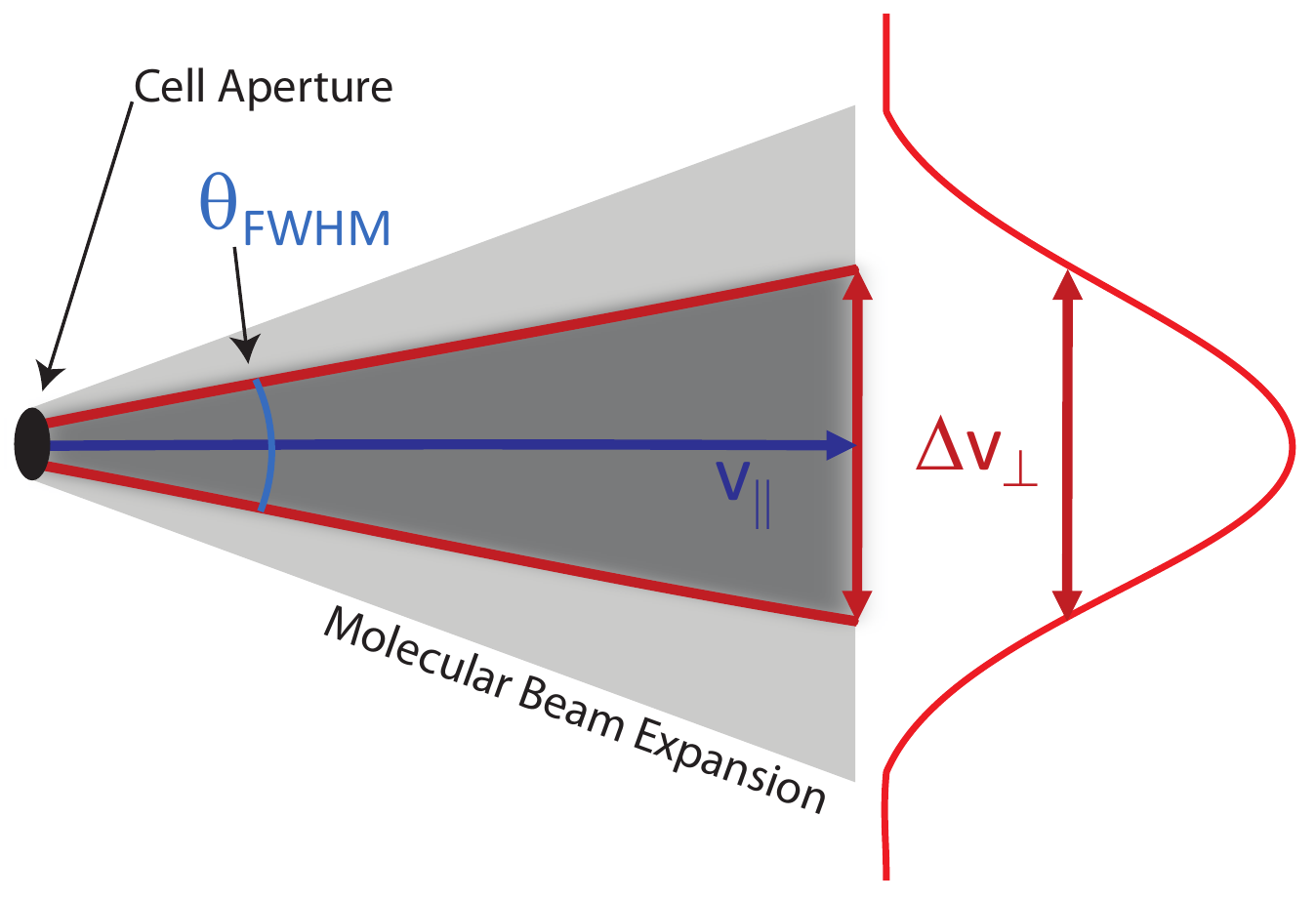}
	\caption{An illustration showing the relationship between the forward velocity $v_{\|}$, transverse velocity spread $\Delta v_{\perp}$, and angular spread $\Delta\theta$ from Eq. (\ref{thetafwhm}), $\tan(\Delta\theta/2) =(\Delta v_\perp/2)/v_\|$.  The gray shaded region indicates the spatial extent of the molecular beam, with darker indicating higher density.}
	\label{fig:angles_schematic}
\end{figure}

We can then also define the solid angle spread $\Delta\Omega$ as the solid angle subtended by $\Delta\theta$, i.e.
\be   \Delta\Omega = 2\pi(1-\cos(\Delta\theta/2)) \ee
Note that this definition ignores the fact that at each point in the beam there is a velocity spread due to the non-zero temperature of the beam, so the measured transverse Doppler spread is a convolution of both beam translational temperature and the actual shape of the beam.  However, if we measure this spread far enough away from the aperture that collisions have stopped, so that the aperture can be regarded as a point source from which the molecules are ballistically expanding, then inferring a spatial spread from the velocity spread is a valid approximation.

For a buffer gas beam in the effusive regime, the angular spread is given by Eq. (\ref{effthetafwhm}) as $\Delta\theta = 120^\circ$, and solid angle spread is given by $\Delta\Omega = \pi$.

In the range $1\lesssim\re\lesssim 10$, the forward velocity begins to increase linearly with $\re$ yet the transverse velocity remains constant at $\Delta v_{\perp,s}=1.5\bar{v}_{0,s}$ from Eq. (\ref{efftransverse}).  Therefore, Eq. (\ref{thetafwhm}) tells us that the divergence will begin to decrease.  As the molecules begin to be boosted to the forward velocity of the buffer gas $v_{\|,s}\approx v_{\|,b}\approx \bar{v}_{0,b}$, the divergence should approach
\bea
\Delta\theta & = & 2\arctan\left(\frac{\Delta v_{\perp,s}/2}{v_{\|,s}}\right) \\
& \approx & 2\arctan\left(\frac{1.5\bar{v}_{0,s}/2}{\bar{v}_{0,b}}\right) \\
& \approx & 2\sqrt{\frac{m_b}{m_s}},
\eea
where in the last approximation we assumed that $m_s\gg m_b$.  The solid angle spread is then approximated by
\be   \Delta\Omega = 2\pi(1-\cos(2\sqrt{m_b/m_s})) \approx \pi m_b/m_s. \ee

Notice that this can be much smaller than the corresponding spreads of $\pi$ for an effusive beam (Eq. (\ref{effomega})), or 1.4 for a supersonic beam (Eq. (\ref{ssomega})).  Both Maxwell et al.\cite{Maxwell2005} (with helium buffer gas) and Hutzler et al.\cite{Hutzler2011} (with neon buffer gas) have operated beams in this regime, and have been able to see the linearly decreasing divergence.

As $\re$ is increased to the point where the transverse spread begins to increase, the divergence will stop decreasing.  Hutzler et al.\cite{Hutzler2011} were able to see this transition for a ThO beam in neon buffer gas, shown in Figure \ref{fig:divergence_vs_flow}.  With helium buffer gas, both Hutzler et al.\cite{Hutzler2011} and Barry et al.\cite{Barry2011} observed that the divergence was constant over a combined range of Reynolds numbers $1\lesssim\re\lesssim 150$.  This was due to the fact that the increases in transverse and forward velocities canceled each other almost exactly.  This indicates that while there is some proposed ``universal shape'' for the relationships of $\re$ vs. $v_\|$ and vs. $\Delta v_\perp$, the Reynolds numbers where transitions in behavior occur for the two relationships need not overlap, and there is not a similar universal shape for $\re$ vs. $\Delta\Omega$. 

\begin{figure}[htbp]
	\centering
		\includegraphics[width=0.45\textwidth]{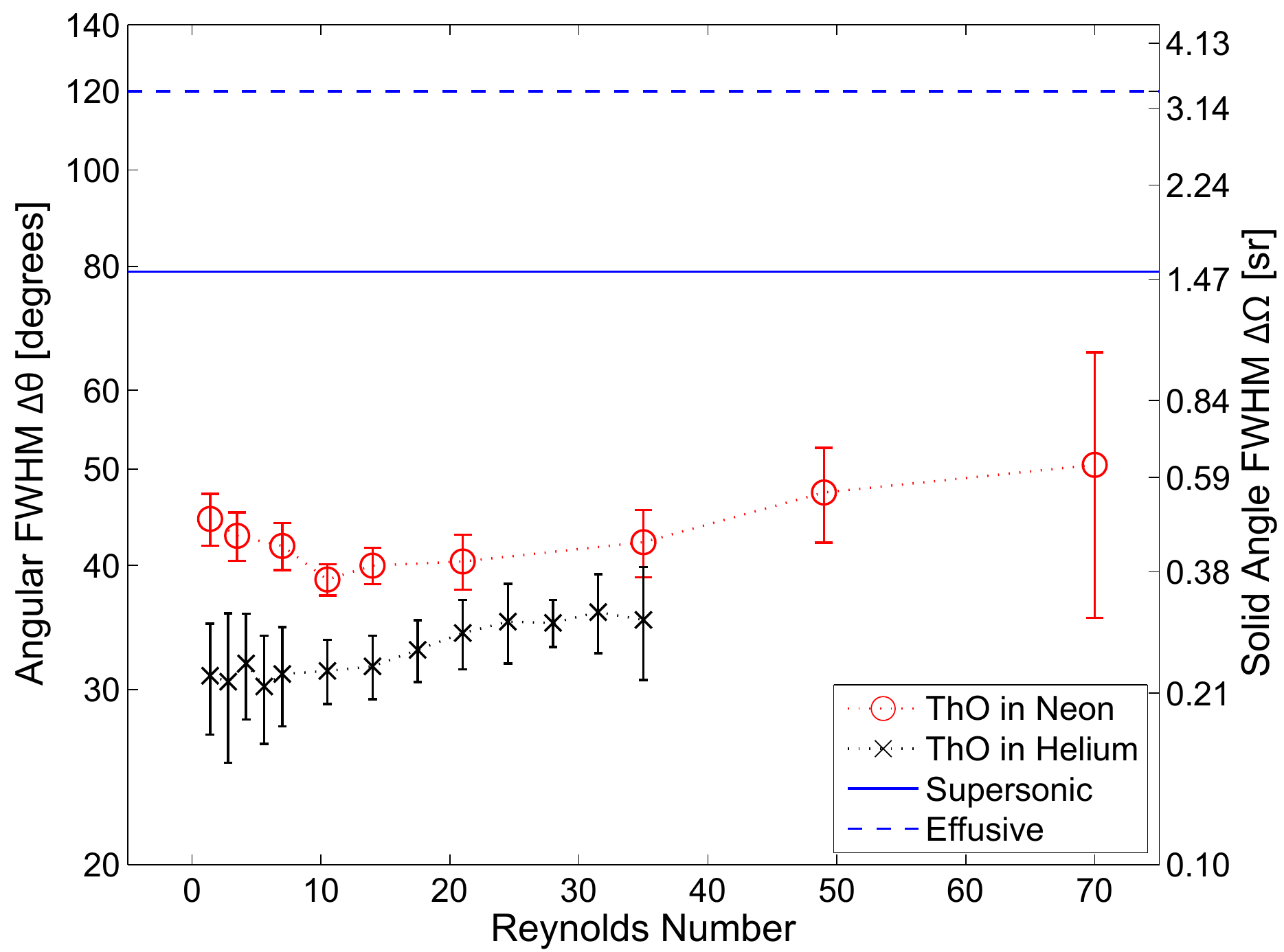}
	\caption{Divergence of a ThO beam in both a helium and neon cooled buffer gas beam\cite{Hutzler2011}.  For low Reynolds number with neon buffer gas, we are in the regime of linearly increasing forward velocity but constant transverse velocity spread (see Figures \ref{fig:vf_vs_flow} and \ref{fig:vt_vs_re}).  Eventually the forward velocity stops increasing and the transverse velocity starts increasing, which results in the divergence ceasing to decrease.  With helium buffer gas, the divergence is essentially constant, similar to what was seen with SrF in a helium buffer gas cooled beam\cite{Barry2011}.  Reproduced from [Hutzler, N. et al., Phys. Chem. Chem. Phys. 13, 18976-18985 (2011) http://dx.doi.org/10.1039/c1cp20901a (2011)] by permission of the PCCP Owner Societies.}
	\label{fig:divergence_vs_flow}
\end{figure}

\subsubsection{Relationship of flow and extraction regimes}\label{sec:revsgammacell}

By examining Eq. (\ref{knre}) for the Reynolds number, which governs the gas flow regime, and Eq. (\ref{gammacell}) for the extraction parameter $\gamma_{cell}$, which governs the species extraction regime, we can see that they are related by a factor which depends on the geometry,
\be \frac{\gamma_{cell}}{\re} \propto \frac{d_{aperture}}{L_{cell}} \label{revsgammacell}.\ee
This means that, at least in principle, it should be possible to design a buffer gas source which is any combination of diffusive (low efficiency) or hydrodynamically extracted (high efficiency), and with effusive, intermediate, or hydrodynamic flow.  Most buffer gas sources operate either in the effusive or intermediate regimes, and it is experimentally challenging to design a beam which is completely effusive, has good extraction, and sufficient thermalization.  A purely effusive beam should have a forward velocity which, according to Eq. (\ref{effvelocity}), does not change with source pressure; however, in buffer gas beam sources with good extraction \cite{Maxwell2005,Hutzler2011,Barry2011,Sommer2009} the forward velocity of the molecules indeed changes with source pressure.  Slowing cells (Section \ref{sec:slowingcells}) can offer near-effusive velocity distributions, but typically have $\sim 1\%$ extraction efficiency\cite{Patterson2007,Lu2011}.

\subsubsection{Choice of buffer gas}\label{sec:bufferchoice}

For many applications, helium is a natural choice of buffer gas.  Helium has large vapor pressure at 4 K (that corresponds to an atom number density of $>10^{19}$ cm$^{-3}$), which is a convenient temperature for cryogenics: Many refrigerators can cool to $\sim 4$ K, as can a simple liquid helium cryostat.  Helium also has the advantage that it has large vapor pressure at even lower temperatures ($< 1$ K), where operating the cell will result in a colder and slower beam\cite{Patterson2007,Patterson2009}.  Simple cryogenic techniques can achieve temperatures of $\sim 2$ K and still handle the necessary heat loads; lower temperatures require a more complicated setup.

Neon has also been used as a buffer gas\cite{Patterson2009,Hutzler2011}, and it has some advantages of its own.  The thermal velocity of the buffer gas scales as $\sqrt{T_0/m_b}$ (Eq. (\ref{vthermal})), so as long as the cell temperature is kept below 20 K with neon buffer gas (whose mass is $\sim 20$ amu), the thermal velocity (and therefore forward velocity; see Section \ref{sec:forwardvelocity}) should be comparable to that of a 4 K helium-4 buffer gas beam source.  Since neon has large enough vapor pressure down to $\sim 14$ K (where the density drops below $\sim 10^{17}$ cm$^{-3}$), it can compare favorably to a 4 K helium source in terms of forward velocity.

\begin{figure}[htbp]
	\centering
		\includegraphics[width=0.40\textwidth]{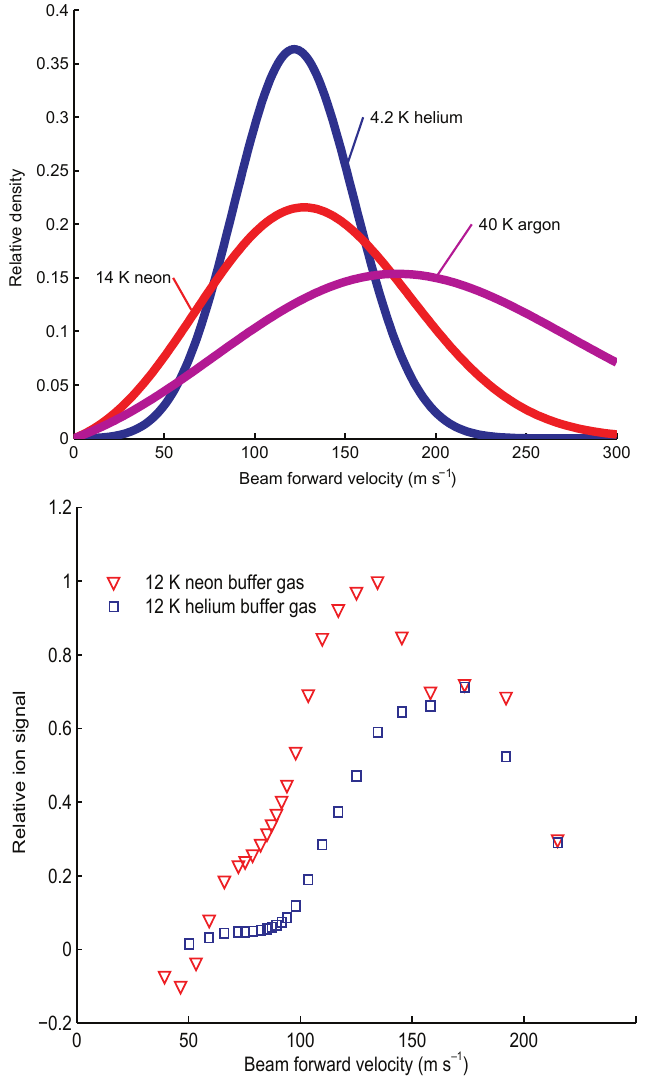}
	\caption{Plots comparing the forward velocity distributions using different buffer gasses\cite{Patterson2009}.  Top: Calculated distributions for noble gases at a temperature where there is sufficient vapor pressure for beam operation.  While helium has sufficient vapor pressure below 4.2 K (and even 1 K), 4.2 K is a typical temperature for both cryogenic refrigerators and liquid helium cooled cryostats.  While the neon distribution is broader (i.e. hotter), for applications where only the slowest molecules are used (such as in trap loading, see Section \ref{sec:collisions}), neon has an advantage.  Bottom: measured velocity distributions of beam of ND$_3$ in helium and neon buffer gas at 12 K.  If the cell must be operated at a higher temperature for technical reasons (for example, if there is a large heat load on the buffer cell from a capillary loading scheme), this data shows that neon has the advantage in terms of creating slow molecules.  The data here is velocity selected.  Reproduced from [Patterson, D., Rasmussen, J., \& Doyle, J. M., New J. Phys. 11(5), 055018 (2009) http://dx.doi.org/10.1088/1367-2630/11/5/055018] with permission from the Institute of Physics.}
	\label{fig:he_vs_ne}
\end{figure}

While the forward velocity of a neon or helium buffer gas cooled beam can be comparable, in certain situations the temperatures in the beam can be comparable as well.  Hutzler et al.\cite{Hutzler2011} found that operating a buffer gas cooled beam resulted in isentropic cooling (see Section \ref{sec:isentropicexpansion}) that reduced the temperature of both a helium and a neon cooled beam of ThO to similar temperatures (see Table \ref{table:beamproperties}), even though the cell temperature was around 17 K for the neon cooled beam and around 4 K for the helium cooled beam.  This means that the beam properties of a helium or neon cooled beam can be similar, with the neon beam offering some distinct technical advantages.  Neon can be efficiently cryopumped by a 4 K surface, while helium requires a large-area adsorbent such as activated charcoal, as discussed in Section \ref{sec:cryopumps}.  Charcoal cryopumps are very effective; however, they must be emptied periodically, have reduced pumping speed as they begin to fill, and can influence the resulting molecular beam properties\cite{Hutzler2011,Barry2011}.  Cryopumping of neon onto a cold surface does not display these properties\cite{Hutzler2011}.  Additionally, the properties of pulsed beams with helium buffer gas can show time-dependence not present with neon\cite{Hutzler2011,Barry2011}.

\subsection{Additional cooling and slowing}\label{sec:additionalcooling}

As discussed in the preceding sections, the properties of a buffer gas beam can be tuned considerably by altering the flow and aperture sizes to control the parameters $\re$ and $\gamma_{cell}$.  However some applications require even finer control; for example, collision studies may require a very narrow and tunable velocity distribution, and trap loading typically requires a very slow distribution (see Section \ref{sec:collisions}).  In this section we shall discuss some methods to manipulate the output of a buffer gas beam, to either reduce temperatures, reduce forward velocities, or select velocity classes.  These techniques help make buffer gas beams highly versatile, and allow the many applications discussed in  Section \ref{sec:applications}.

\subsubsection{Isentropic expansion}\label{sec:isentropicexpansion}

The boosting effect discussed in Section \ref{sec:forwardvelocity} can be detrimental if one aims to produce a slow beam, which is one of the benefits of buffer gas beams over supersonic beams and effusive ``oven'' beams.  One would therefore prefer to operate the beam in the effusive flow regime, i.e. at a low Reynolds number (Eq. (\ref{knre})).  It is possible to keep the Reynolds number fairly low while maintaining good extraction $(\gamma_{cell}>1)$ by using a slit-shaped aperture, as we will now discuss.  Consider the aperture as having a short and long dimension, so $A_{aperture}=d_{short}\times d_{long}$.  For a fixed internal cell geometry, buffer gas, species, and cell temperature, we can see from Eqs. (\ref{knre}) and (\ref{gammacell}) that
\bea \re & \propto & d_{short} n_{0,b}\\
\gamma_{cell} & \propto & n_{0,b} A_{aperture} \\
\Rightarrow \frac{\gamma_{cell}}{\re} & \propto & d_{long} \eea
Therefore, by increasing $d_{long}$ but keeping $A_{aperture}$ fixed, we can decrease $\re$ without changing $\gamma_{cell}$ or the buffer gas density.  Note that simply changing the aperture size while leaving all other parameters fixed also has the effect of changing $\re$ but not $\gamma_{cell}$ (Eq. (\ref{gammacell})); however, this is not ideal for two reasons: First, the in-cell buffer gas density is constrained to be large enough that the species is thermalized, but small enough that the molecules can diffuse away from the injection point\cite{Skoff2011}, and this method would change the buffer gas density.  Second, as mentioned in Section \ref{sec:extraction}, reducing the aperture size can in fact have an effect on the extraction for small enough aperture sizes.  For these reasons, earlier buffer gas beam papers tended to use a slit\cite{Maxwell2005} aperture, typically around $1\times 5$ mm.

While performing buffer gas beam studies that required high flow rates to sufficiently thermalize atoms from a hot oven,\cite{Patterson2009,Lu2009} it was quickly realized that the downside of high $\re$ flows could also come with a benefit: isentropic cooling from the free expansion of a gas, similar to what happens in supersonic beams\cite{Scoles1988}.  This effect was first seen (though not published) in K and Rb atoms cooled by a neon buffer gas\cite{Patterson2009,Lu2009}, and it was subsequently fully characterized with ThO in neon and helium buffer gas beams at cell temperatures of about 17 K and 4 K (respectively)\cite{Hutzler2011}, and simultaneously with SrF in helium buffer gas at a cell temperature of 3 K\cite{Barry2011}.  In each case, the molecules were found to cool rotationally and translationally below the temperature of the cell, and the beam properties had very good qualitative agreement.  Selected properties of these beams are listed in Table \ref{table:beamproperties}.

The cell apertures for these studies were either round or square, and the studies were performed between $\re\approx 1-150$.  In each case, the Reynolds numbers were high enough that the rotational temperatures appeared to saturate around 2 K and 1 K for ThO and SrF respectively, as shown in Figure \ref{fig:t_rot_fwd}.  The translational temperatures (in the forward direction) of ThO in neon buffer gas was also decreased below the temperature of the cell, and is shown in Figure \ref{fig:t_rot_fwd}.  However, the translational temperatures for both ThO and SrF in helium buffer gas were complicated by time-dependence of the temperature after the ablation pulse, though in each case it was reduced below the cell temperature.

It is worth noting that even in these fully-boosted, high Reynolds number regimes, where the forward velocity is fully saturated (see Section \ref{sec:forwardvelocity}), the velocity still compares very favorably to supersonic or effusive beams (see Tables \ref{table:beamcomparison} and \ref{table:beamproperties}).  In the next section, we shall see that there are additional techniques to further reduce the forward velocity of a buffer gas cooled beam.

\begin{figure}[htbp]
	\centering
		\includegraphics[width=0.45\textwidth]{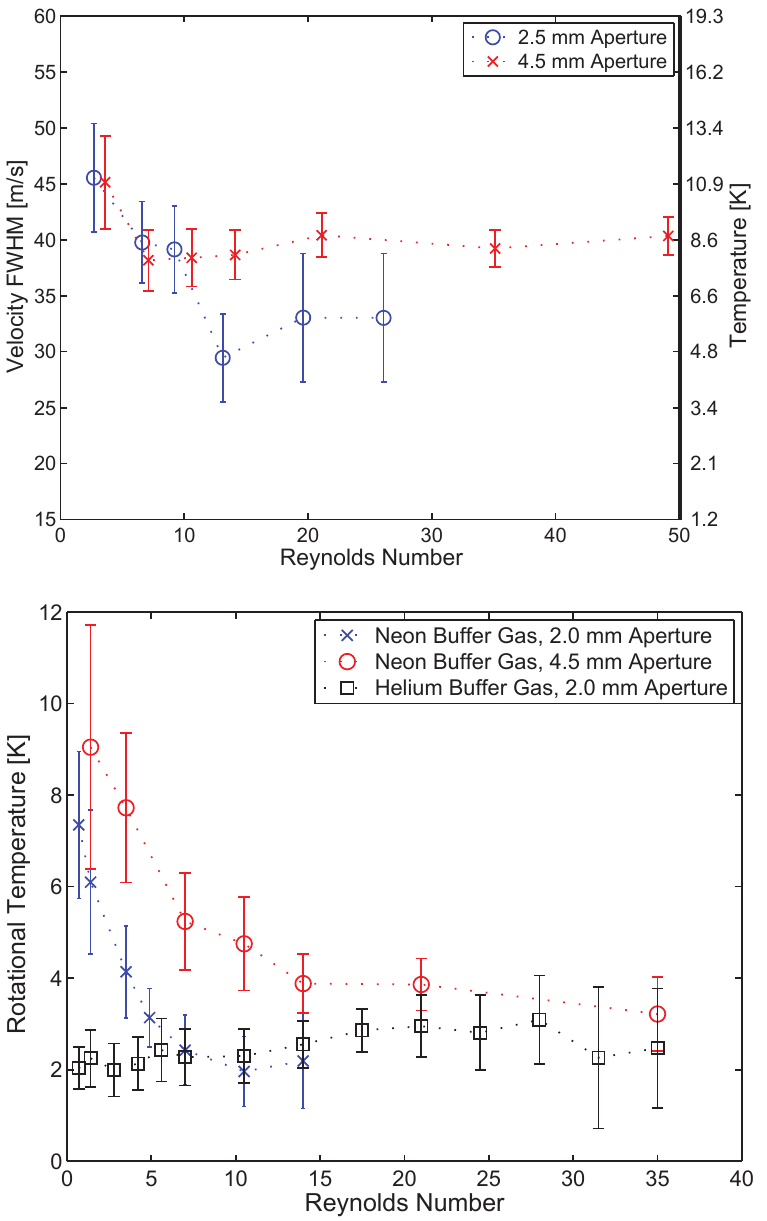}
	\caption{Additional cooling from isentropic expansion in a buffer gas cooled ThO beam\cite{Hutzler2011}.  The cell temperature is around 17 K for neon buffer gas, and around 4 K for helium buffer gas.    Top: Longitudinal velocity spread (temperature) in a neon cooled buffer gas beam, with two different aperture sizes.  Bottom: Final rotational temperature in the beam, for comparing two different aperture sizes with neon buffer gas, and a single aperture size for helium buffer gas.  The rotational temperature with helium buffer gas showed essentially no dependence on the aperture size.  Figures are reproduced from [Hutzler, N. et al., Phys. Chem. Chem. Phys. 13, 18976-18985 (2011) http://dx.doi.org/10.1039/c1cp20901a (2011)] by permission of the PCCP Owner Societies.}
	\label{fig:t_rot_fwd}
\end{figure}

\subsubsection{Slowing cells}\label{sec:slowingcells}

As mentioned in the previous section, the beam properties can be manipulated by choice of aperture geometry.  This control can be extended by using an advanced aperture geometry known as a ``slowing cell''\cite{Patterson2007,Lu2011}, shown in Figure \ref{fig:slowing_cell}, which is used to create a region of intermediate pressure between the cell and vacuum (beam) regions, and acts as an effusive source that can still utilize the high extraction of the main cell.

\begin{figure}[htbp]
	\centering
		\includegraphics[width=0.45\textwidth]{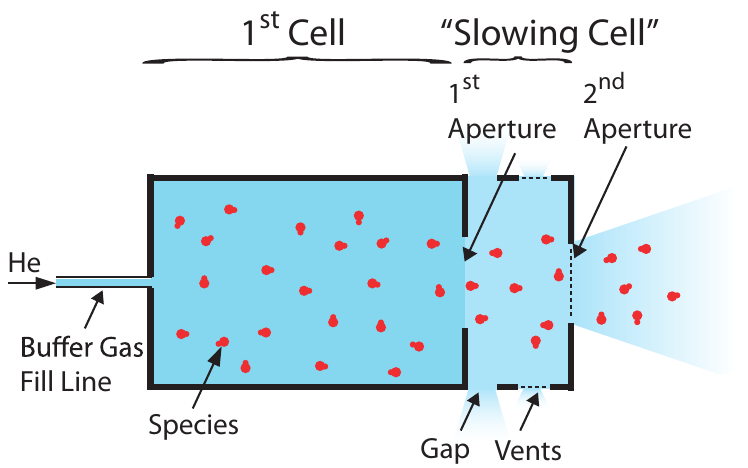}
	\caption{A slowing cell, adapted from Lu et al.\cite{Lu2011}.  Here the species is represented by red dots, and the helium buffer gas is represented by blue shading, where darker shading indicates higher density.  The density in the slowing cell is kept low by gaps and vents, so that the species suffers only a few collisions in the slowing cell.  These collisions can reduce the forward velocity without heating, while sacrificing flux.  The method for introducing the species is not indicated in this figure, but has been accomplished with both ablation\cite{Patterson2007,Lu2011} and capillary filling\cite{Patterson2007}.}
	\label{fig:slowing_cell}
\end{figure}

The main, or ``first,'' cell is created like a typical buffer gas cell, and has the same typical buffer gas flow rates, dimensions, etc.  A second, or slowing, cell is then attached to the aperture of the main cell.  This second cell has a very large area through which the buffer gas can flow, so the steady-state pressure inside it is low, about 10\% of the pressure in the main cell.  This makes the typical mean free path on the order of a few mm, so the species will scatter only a few times while in the slowing cell.  If the slowing cell is also kept cold, these collisions will not heat the molecules, but will have the effect of reducing their boosted forward velocity closer to the thermal velocity (see Section \ref{sec:forwardvelocity}).

\begin{table}[htbp]
	\centering
		\begin{tabular}{llllllll}
			Species & $T_0$ [K] & $v_\|$ [m s$^{-1}$] & $N$ [pulse$^{-1}$] \\ \hline
			Yb, Boosted\cite{Patterson2007}  & 2.6 & 130 & $5\times 10^{12}$ \\
			Yb, Slowed\cite{Patterson2007}  & 2.6 & 35 & $5\times 10^{10}$ \\
			CaH, Boosted\cite{Lu2011} & 1.8 & 110 & $5\times 10^{9}$ \\
			CaH, Slowed\cite{Lu2011} & 1.8 & 65 & $1\times 10^{9}$ \\
			CaH, Slowed\cite{Lu2011} & 1.8 & 40 & $5\times 10^{8}$ \\
		\end{tabular}
	\caption{A comparison of boosted buffer gas beam sources (i.e. with no slowing cell), to those with a slowing cell.  ``Boosted'' means that collisions near the aperture accelerate the beam above the thermal velocity, as discussed in Section \ref{sec:forwardvelocity}.  ``Slowed'' means a slowing cell was added to the main cell.}
	\label{table:slowingcell}
\end{table}

Notice that operating a beam in the effusive regime can also result in low forward velocities (see Section \ref{sec:forwardvelocity}); however, as mentioned earlier, it is difficult to operate a fully effusive beam that also has significant extraction.  The PbO beam of Maxwell et al.\cite{Maxwell2005} had a forward velocity of $\sim 40$ m s$^{-1}$ at low flows, comparable to those achieved with Yb and CaH using slowing cells, but with an extraction fraction of $f_{ext}<10^{-4}$, compared to $f_{ext}\approx 10^{-2}$ for the slowing cells.

\subsubsection{Guiding and velocity filtering}\label{sec:guiding}

Although they may not technically constitute cooling or slowing, electric or magnetic guides or velocity filters can be used to change the output of a buffer gas cooled beam to suit a particular application\cite{Patterson2007,Patterson2009,Sommer2009,Buuren2009,Sommer2010}.  Guiding the species of interest can be useful for a number of reasons.  First, if the presence of the buffer gas will interfere with the experiment, for example while performing collision studies, a guide can be used to transport the species far from the buffer gas source where there is no appreciable buffer gas pressure\cite{Sawyer2011}.  Second, if the beam is to be studied in a room temperature apparatus, for example to perform laser cooling \cite{Shuman2010} or precision measurements\cite{Vutha2010}, the beam may have to travel a long distance to exit the cryogenic apparatus before passing into the experimental region.  Without a guide, the molecule density will decrease with the square of the distance from the cell, which could result in significant reduction.  An electrostatic or magnetic guide or lens can help reduce this loss from beam divergence.

Magnetic $\vec{\mathcal{B}}$ or electric $\vec{\mathcal{E}}$ fields can be used to guide or focus a molecular beam by creating a restoring force perpendicular to the motion of the molecule\cite{Scoles1988}.  If the molecular state has a magnetic moment $\mu_\mathcal{B}$, then the interaction (of a low-field seeking state) with a magnetic field leads to a potential energy $U = \mu_\mathcal{B} |\vec{\mathcal{B}}|$.  If we assume that the molecule is traveling in a cylindrically symmetric field $\vec{\mathcal{B}}(r)$, where $r$ is the distance from the axis of symmetry, then the field from a $n-$pole configuration is $|\vec{\mathcal{B}}(r)| = \mathcal{B}_0 r^{(n-2)/2}$, and the potential energy of the molecule is $U = \mu_\mathcal{B} \mathcal{B}_0 r^{(n-2)/2}$.  In addition to guiding, multipole fields can also be used to focus molecules by creating a linear restoring force $(U\propto r^2)$, which we can see requires a hexapole $(n=6)$ field configuration.

The situation is similar for polar molecules.  If a molecular state has a permanent electric dipole moment $\mu_\mathcal{E}$ (as is the case for states with $\Lambda>0$, such as $\Pi$ or $\Delta$ states\cite{Herzberg1989}), then the potential energy of the (low-field seeking state) molecule in an $n-$pole field is $U = \mu_\mathcal{E} \mathcal{E}_0 r^{(n-2)/2}$.  However, unlike the case with magnetic dipole moments, polar molecules can be in states (such as $\Sigma$ states\cite{Herzberg1989}) which have an induced dipole moments, leading to a quadratic shift in fields.  In this case, the dipole moment (for small fields) is proportional to the electric field $\mu_\mathcal{E}\propto |\vec{\mathcal{E}}|$, so potential energy of the molecule is $-\mu_\mathcal{E}|\vec{\mathcal{E}}| \propto r^{n-2}$.  If we wish to focus polar molecules with a linear restoring force, we can see that a hexapole is required for states with a permanent dipole moment, or a quadrupole $(n=4)$ for states with an induced dipole moment.

If the maximum electric field in the guide (typically $\sim10-40$ kV cm$^{-1}$) is $\mathcal{E}_{max}$ and the molecules have Stark shift $W(\mathcal{E})$ in a field $\mathcal{E}$, then the transverse depth of the guide is defined as $W_{max} = W(\mathcal{E}_{max})$, which is typically $\sim 1$ K$\times k_B$.  This sets the maximum transverse velocity spread in the guide $v_{\perp,max}$ to be
\be \frac{1}{2}m_s v_{\perp,max}^2 = W_{max} \Rightarrow v_{\perp,max} = \sqrt{2W_{max}/m_s}, \ee
which is around 10-30 m s$^{-1}$.

\begin{figure}[htbp]
	\centering
		\includegraphics[width=0.45\textwidth]{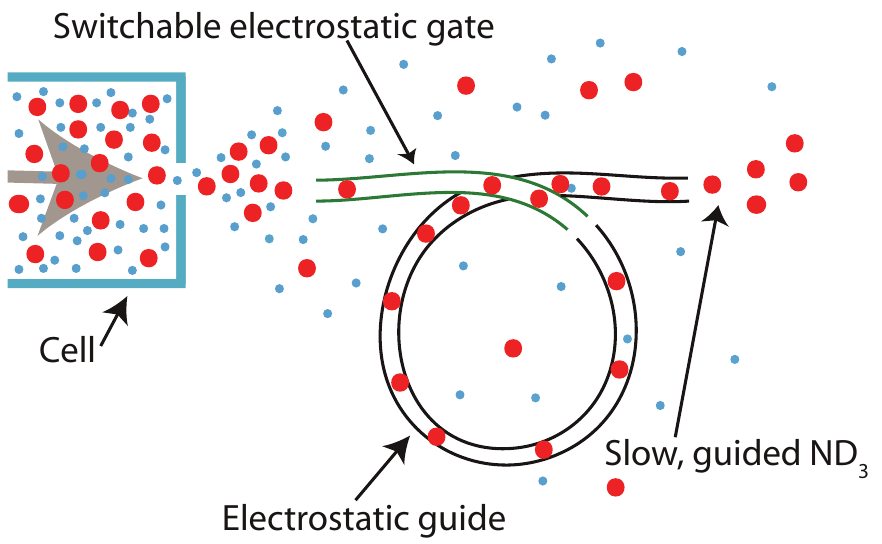}
	\caption{Electrostatic guide for ND$_3$\cite{Patterson2009}.  The cell is kept at 15 K and contains neon buffer gas (small blue circles) and ND$_3$ (large red circles) introduced from a capillary.  As the molecules exit the cell, an ``electrostatic gate'' can be turned on to direct molecules into the guide.  This guide will only admit slow molecules around the bend; faster molecules will simply overcome the potential barrier and escape the guide. Adapted from Patterson et al.\cite{Patterson2009}}
	\label{fig:nd3_guide}
\end{figure}

In addition to guiding or focusing to increase useful flux, electric or magnetic fields can be used to select particular velocity ranges from the output of the molecular beam, which is useful for collision studies and trap loading experiments (see Section \ref{sec:collisions}), as will be discussed.  Molecules of a particular forward velocity range can be selected by rotating mechanical objects \cite{Szewc2010}; however, using fields is much less technically challenging, especially in a cryogenic environment.  The first realization of this method\cite{Rangwala2003} (and later realizations with buffer gas cooled beams\cite{Buuren2009,Patterson2007,Patterson2009,Sommer2009}) was to simply bend one of the guides described above, similar to what is shown in Figure \ref{fig:nd3_guide}.  Molecules with high enough kinetic energies in the forward direction will simply pass over the potential barrier, and only the more slow-moving molecules will remain in the guide.  More specifically, for a guide of depth $W_{max}$, guide radius $\rho$ (i.e. distance from the center of the guide to the location of the maximum electric field), and guide radius of curvature $R$, the cutoff for longitudinal velocities which will remain in the guide is found\cite{Sommer2010} by equating the centrifugal force $m_s v_{\|,max}^2/R$ to the restoring force $\sim W_{max}/\rho$ caused by the molecule traveling up the potential hill of length $\rho$, or
\be m_s v_{\|,max}^2/R \approx W_{max}/\rho \Rightarrow v_{\|,max} \approx \sqrt{W_{max} R/\rho m_s} \ee
This technique can be used to filter only slow molecules from a pulsed or continuous buffer gas source, for example to load a trap.  In fact, this technique can also be applied to thermal effusive sources; Rangwala et al.\cite{Rangwala2003} filtered slow H$_2$CO and ND$_3$ molecules from a 300 K effusive source and obtained distributions that had widths and means similar to those of a 5 K thermal distribution, while maintaining a high flux of $\approx 10^9$ s$^{-1}$ guided H$_2$CO molecules.

In addition to simply selecting slow molecules, electrostatic guides can be used to select a window of velocities by using multiple switched guides.  Sommer et al.\cite{Sommer2010} devised a three-stage, switchable electrostatic guide, fed by a helium buffer gas cooled source of ND$_3$ molecules, which could deliver controllable, velocity-selected molecule pulses.  By switching the individual guide segments on or off at precise times, both a lower and upper velocity cutoff can be imposed on molecules that remain in the guide.  Velocity windows of FWHM $\sim 3-20$ m s$^{-1}$ wide, with centers between 20 and 100 m s$^{-1}$, could be efficiently selected with low losses, as shown in Figure \ref{fig:velocity_selection}.

\begin{figure}[htbp]
	\centering
		\includegraphics[width=0.45\textwidth]{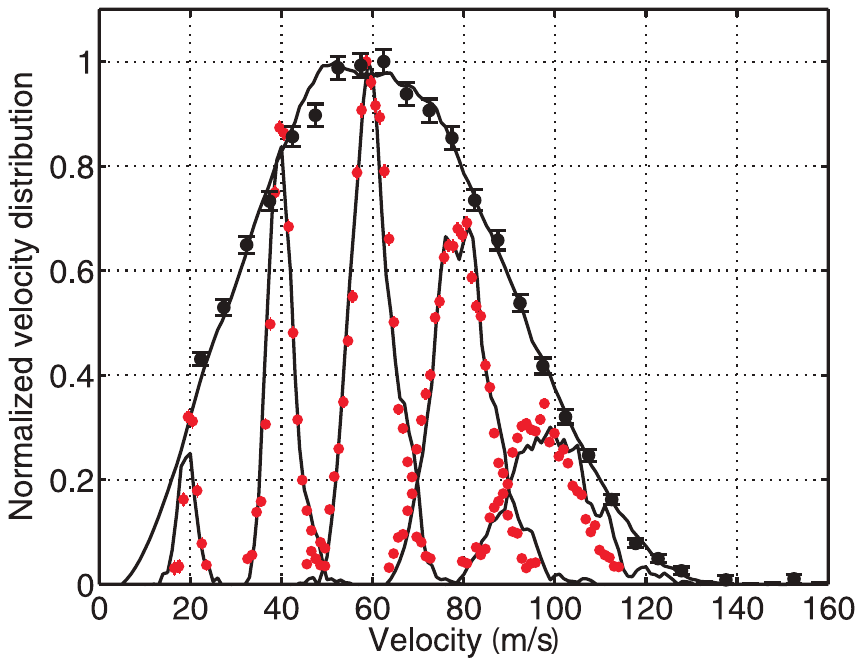}
	\caption{Velocity-selected pulses of ND$_3$ obtained using multiple bent, switchable guides\cite{Sommer2010}.  The pulse centered about 60 m s$^{-1}$ contains approximately 10$^5$ molecules.  Reproduced from [Sommer, C. et al., Phys. Rev. A 82(1), 013410 (2010) http://dx.doi.org/10.1103/PhysRevA.82.013410].  Copyright (2010) by the American Physical Society.}
	\label{fig:velocity_selection}
\end{figure}

\subsection{Effusive and supersonic beam properties}

For the sake of comparison, we will briefly review some properties of effusive and supersonic beams.  Detailed discussions may be found in existing literature\cite{Scoles1988,Pauly2000}.

\subsubsection{Effusive beams}\label{sec:effusive}

In an effusive gas flow from an aperture, the typical escaping gas atom has no collisions near the aperture ($\kn > 1$).  Therefore, the resulting beam can be considered as a sampling of the velocity distribution in the cell.  A typical setup is a gas cell with a thin exit aperture, where the thickness of the aperture and the aperture diameter $d_{aperture}$ are both much smaller than the mean free path of the gas at stagnation conditions.  Note that the effusive beams under discussion in this section are oven-type effusive sources, i.e. where the vapor pressure of the species of interest is large at the source temperature (as opposed to buffer gas cooled effusive sources, which can operate at temperatures where the species has no appreciable vapor pressure).

The number density in the beam resulting from a differential aperture area $dA$ is given by
\bea
n(R,v,\theta) & = & n(R,\theta)f(v), \textrm{ where} \\
n(R,\theta) & = & \frac{n_0\cos(\theta)}{4\pi R^2}dA, \\
f(v) & = & \frac{32}{\pi^2\bar{v}^3}v^2 e^{-4 v^2/\pi\left(\bar{v}\right)^2},
\eea
$R$ is the distance from the aperture, $\theta$ is the angle from the aperture normal, $n_0$ is the stagnation density in the cell, $n(R,\theta)$ is the total number density distribution integrated over velocity, and $f(v)$ is the normalized velocity distribution in the cell.  The velocity distribution in the beam is given by \cite{Pauly2000}
\be
f_{beam}(v) = (v/\bar{v}) f(v) = \frac{32}{\pi^2\bar{v}^4}v^3 e^{-4 v^2/\pi\left(\bar{v}\right)^2}.
\ee

From the velocity distribution we can extract the mean forward velocity of the beam,
\be
\bar{v}_{\|,ef\! f} = \int_0^\infty v f_{beam}(v)\;dv = \frac{3\pi}{8}\bar{v} \approx 1.2 \bar{v}
\label{effvelocity}\ee

From the total number density in the beam $n(R,\theta)$, we can extract the FWHM (full-width at half-maximum) of the characteristic angular spread $\Delta\theta$ by solving $n(R,\Delta\theta/2) = \frac{1}{2}n(R,0)$, which gives
\be
\Delta\theta = \frac{2\pi}{3} = 120^\circ,
\label{effthetafwhm}\ee
or the characteristic solid angle $\Delta\Omega$
\be
\Delta\Omega = 2\pi(1-\cos(\Delta\theta/2)) = \pi \label{effomega}
\ee
Note that this is the angular spread from a differential area of the aperture; near the aperture we would have to integrate over the area of the aperture to obtain the full shape, however this angular spread is valid in the far field.

The rate at which molecules escape the cell and therefore enter the beam is simply the molecular flow rate through an aperture,
\be   \dot{N} = \frac{1}{4}n_{0,b}\bar{v}_0 A_{aperture}. \ee

Effusive beams of certain species have very large fluxes, such as certain metal atoms or low-reactivity molecules with high vapor pressure.  For alkali metals \cite{Metcalf1999}, the cell (or oven) temperature required to achieve a vapor pressure of 1 torr is between around 500 and 1000 K.  The flow through a 1 mm$^2$ aperture is then around $10^{14}-10^{15}$ s$^{-1}$.  Note that in this situation there is no buffer gas.

The beam resulting from an effusive source is not immediately useful for many applications.  The large forward velocity (typically several hundred m s$^{-1}$) would limit laser interrogation time, and broad velocity distributions would lead to significantly broadened spectral lines.  Some atoms can be slowed and cooled using powerful optical techniques\cite{Metcalf1999}; however, until recently molecules have (with few exceptions \cite{Shuman2009,Shuman2010}) resisted these optical techniques due to their complex internal structure.  In addition, molecules have internal degrees of freedom which can be excited by the large temperatures in an oven.  The rotational energy constant for diatomic molecules is typically 1-10 K $\times k_B$, so at typical oven temperatures the molecules can be distributed over hundreds of rotational states.

\subsubsection{Fully hydrodynamic, or ``supersonic'' beams}\label{sec:supersonic}

In a fully hydrodynamic, or ``supersonic'' beam, the gas experiences many collisions near the exit aperture $\kn\ll 1$ (typically $\kn<10^{-3}$), and therefore the beam properties are determined by the flow properties of the gas.  In this case we cannot apply simple gas kinetics as in the case of effusive beams, but instead must consider the dymanics of a compressible fluid.  We shall assume that the gas is monoatomic.

A typical supersonic source has $P_0\sim 1$ atm, and $d_{aperture}\sim 1$ mm$^2$.  Supersonic sources are often cooled so that the forward velocity is reduced, and typically have $T_0=200 - 300$ K.  Increasing the backing pressure $P_0$ can be used to further reduce the temperature of the molecules, though at high enough pressures cluster formation can begin to negatively affect the beam properties\cite{Scoles1988}.  On the other hand, buffer gas beams can be operated with high enough backing pressure such that increasing the pressure further will no longer reduce the beam temperature, without degradation of the beam properties\cite{Hutzler2011,Barry2011}.  Additionally, for supersonic beams, the introduction of chemically reactive or refractory species is challenging due to the short mean free path between collisions in the source, and are typically introduced into the expansion plume\cite{Hopkins1983,Fletcher1993,Dietz1981,Tarbutt2002}.  This reduces the brightness of the beam.  On the other hand, beams of polar radicals can be created with buffer gas cooling\cite{Hutzler2011,Barry2011} with around 100 times the brightness of supersonic beams of similar species (see Table \ref{table:beamcomparison}).

The gas flow rate from the aperture in a supersonic source can be on the order of 1 standard liter per second, which would make keeping good vacuum in the apparatus difficult if the beam were to operate continuously.  For this reason, supersonic beams are often pulsed to reduce time-averaged gas load on the vacuum system.  Continuous, or ``Campargue''-type\cite{Campargue1984,Scoles1988} supersonic beams are possible, although they introduce many technical challenges.  On the other hand, buffer gas cooled beams can be operated continuously without considerable difficulty (see Table \ref{table:beamproperties} for a list of continuous buffer gas cooled beams).

For a monoatomic gas with specific heat ratio $\gamma=5/3$, the number density and temperature in a supersonic expansion are related by\cite{Pauly2000}

\be   \frac{n}{n_0} = \left(\frac{T}{T_0}\right)^{3/2}. \label{thermoscaling}\ee

In the far field (typically more than four times the aperture diameter away from the aperture \cite{Pauly2000}), the number density will fall off as a point source, $n(R)\propto R^{-2}$, where $R$ is the distance to the aperture.  Therefore,
\be   n(R) \propto R^{-2}, \quad T(R) \propto R^{-4/3}. \ee

Thus we see that, unlike in the case of an effusive beam, as the beam expands the temperature decreases.  The temperature will continue to decrease until the gas density becomes low enough that collisions stop, the gas ceases to act like a fluid, and the gas atoms simply fly ballistically.  This transition is often called ``freezing'' or ``quitting.''  Even from a room-temperature supersonic source, it is not uncommon to have beam-frame temperatures of around 1 K.

The relationship between the forward velocity and temperature of an ideal monoatomic gas expansion is given by \cite{Pauly2000}
\be   v_\| = \sqrt{\frac{5 k_B (T_0-T)}{m}}. \label{vfvstemp}\ee
If this gas is allowed to expand a long enough distance such that $T\ll T_0$, then the final forward velocity is
\be   v_\| = \sqrt{\frac{5 k_B T_0}{m}} = \bar{v}_0\sqrt{\frac{5\pi}{8}}  \approx 1.4 \bar{v}_0. \label{finalvf} \ee
A standard supersonic source is argon expanding from a 300 K cell, which has a forward velocity of about 600 m s$^{-1}$.  A more technically challenging source has argon expanding from a 210 K cell, which has a forward velocity of about 300 m s$^{-1}$.  On the other hand, buffer gas beams typically have a forward velocity of $<200$ m s$^{-1}$, and can be as slow as 40 m s$^{-1}$ (see section Section \ref{sec:forwardvelocity}).

The label ``supersonic'' comes from the fact that as the gas expands from the aperture, the temperature drops so that the speed of sound (Eq. (\ref{speedofsound})) decreases yet the forward (flow) velocity increases (Eq. (\ref{vfvstemp})), so that eventually the Mach number becomes larger than 1.  In fact, setting the stagnation pressure and temperature so that the Mach number is exactly 1 at the aperture yields maximum gas flow rate through the aperture \cite{Pauly2000}.

The number density as a function of the distance $R$ and the angle $\theta$ from the aperture normal is given approximately by \cite{Ashkenas1966}
\be   n(R,\theta) = n(R,0)\cos^2\left(\frac{\pi\theta}{2\phi}\right), \ee
where $\phi \approx 1.4$ for a monoatomic gas.  We can then find the angular spread $\Delta\theta$ and solid angle $\Delta\Omega$ as in Eqs. (\ref{effthetafwhm}) and (\ref{effomega}) to be
\be   \Delta\theta = \phi \approx 1.4 = 79^\circ, \quad \Delta\Omega \approx 1.4 \label{ssomega}\ee

If there is a small amount of a species of interest mixed in with the main (or carrier) gas, there are enough collisions that the species will follow the carrier gas flow lines, and always be in thermal equilibrium.  Therefore the properties discussed for the carrier gas should be very similar for the species gas as well.  If the species has internal structure (i.e. electonic, vibrational, and rotational states that can be thermally populated), then if there are sufficiently large inelastic, internal-state-changing collision cross sections with the carrier gas then the internal temperatures can be thermalized as well\cite{Scoles1988}.  This is often the case, so supersonic beams are useful tools for creating beams of translationally and internally cold molecular beams, albeit with a large forward velocity.

\subsection{Technical details of source construction}\label{sec:technicaldetails}

The basic requirements for a buffer gas cell were outlined in Section \ref{sec:cell}, but in this section we will provide some technical details of source construction.  More details may be found in existing literature \cite{Maxwell2005,Patterson2007,Patterson2009,Hutzler2011,Barry2011}.  Here we shall focus purely on the design and engineering aspects, so readers not concerned with these details are encouraged to skip this section.

\subsubsection{The cell}

Here we will describe the cell used by Hutzler et al.\cite{Hutzler2011}, however the features are similar for other ablation-loaded buffer gas cells\cite{Maxwell2005,Patterson2007,Barry2011} (we will not consider the technical details of capillary loading\cite{Patterson2007,Patterson2009}). A schematic of the cell is shown in Figure \ref{fig:buffergasbeam}.

  The cell is a cryogenically cooled, cylindrical copper cell with internal dimensions of 13 mm diameter and 75 mm length.  A 2 mm inner diameter copper tube entering on one end of the cylinder flows buffer gas into the cell. A 150 mm length of the fill line tube is thermally anchored to the cell, ensuring that the buffer gas is cold before it flows into the cell volume.  An open aperture (or nozzle) on the other end of the cell lets the buffer gas spray out as a beam.  The aperture should be thin-walled (typically <0.5 mm) to prevent the species of interest from sticking to the sides of the aperture.  The ablation target is located approximately 50 mm from the exit aperture. A pulsed Nd:YAG laser (Continuum Minilite II, 532 nm, 17 mJ per 6 ns pulse), focused with a converging lens, is fired at the target.

In addition to the main, cylindrical internal volume, two holes are drilled perpendicularly to the main volume axis.  One hole (13 mm diameter) has the ablation target on one side, and a window on the other; another hole (10 mm diameter) has a window on both sides for a spectroscopy laser.  The windows are made of 3 mm thick borosilicate glass, and are sealed to the holes in the cell with indium.  The window for the ablation laser is mounted on the end of a 38 mm long copper tube which attaches to the cell; this keeps the window far from the ablation target.  If the window is too close to the ablation target, the window will become coated with ablation detritus, into which the ablation laser can begin depositing energy.  This both reduces the amount of energy deposited into the ablation target, and cause the window to become damaged and unusable.

\subsubsection{Radiation shields}
The blackbody heat load on a 4 K surface from a 300 K environment is approximately 50 mW cm$^{-2}$.  A cryogenic cell can easily have surface area of $>$100 cm$^2$, and would therefore have to experience a heat load of several watts.  This would overwhelm most cryogenic refrigerators, so it is necessary to surround the cell with cold surfaces.  Surrounding the cell by a blackbody shield cooled by a liquid nitrogen cryostat (77 K), or the first stage of a pulse tube cooler ($\sim 30-50$ K) will reduce the blackbody heat load on the cell to $<1\%$ of the 300 K blackbody heat load.  This heat load will instead be deposited into the blackbody shield; however, a typical liquid nitrogen cryostat or pulse tube cooler can easily handle these heat loads at these temperatures.  The blackbody heat load can also be further reduced by covering the blackbody shield with several layers of thin aluminum-coated mylar ``superinsulation''.

The cell is also sometimes additionally enclosed in an inner radiation shield, typically around 4-5 K.  This can help reduce the heat load further, since it could be easier to superinsulate a radiation shield instead of the cell itself.  Additionally, the 4 K shield could help to protect cryopumps, as discussed below.

\subsubsection{Cryopumps}\label{sec:cryopumps}
Enclosing the cell in radiation shields means that the gas conductance to the external vacuum chamber is very low.  If no pumping occurred inside the radiation shields, the buffer gas would build up to the point where the residual pressure was too high to allow for a beam.  In the case of neon buffer gas, this can be solved by surrounding the cell with a radiation shield kept at a temperature where the vapor pressure of neon is acceptably low, which is simple to achieve with standard cryogenic refrigerators or liquid helium (for example, the vapor pressure of neon is $<10^{-9}$ torr at 7 K).  A neon cooled buffer gas beam inside a 4 K radiation shield can operate for extended periods of time; Hutzler et al.\cite{Hutzler2011} operated such a beam continuously for over 24 hours without beam or vacuum degradation.

For helium this is not practical; the radiation shield would have to be cooled to $<0.5$ K for similar vapor pressures, which would require much more complex cryogenic systems and superfluid film management.  Fortunately there is a solution, which is to use activated charcoal as a cryopump\cite{Pobell1996}.  When cooled to $\lesssim 10$ K, activated charcoal becomes a cryopump for helium with up to several l s$^{-1}$ pumping speed per cm$^2$ of charcoal, and can hold almost 1 STP liter of helium per gram (though these values are highly dependent on temperature, and other parameters \cite{Pobell1996}).   Covering the inside of a 4 K radiation shield with $\sim 100$ cm$^2$ of activated charcoal bonded to copper plates can result in a cryopump with a pumping speed of several hundred liters per second that can last for hours under typical gas loads\cite{Hutzler2011,Barry2011}.  As the charcoal fills up with helium the pumping speed begins to change, which can result in a degradation of beam properties \cite{Hutzler2011,Barry2011}; therefore, it is necessary to warm up the charcoal (typically to $\sim20$ K) and pump out the desorbed helium with a mechanical pump.  Experience shows that the beam properties are very sensitive to charcoal amount and placement\cite{Hutzler2011,Barry2011}, often in very non-intuitive ways.

\subsubsection{Beam collimation}  Similar to supersonic beams, buffer gas cooled beams typically have collimators to create differentially pumped regions and reduce the gas load on the part of the apparatus where the molecules are being probed.  With buffer gas cooled beams this is especially advantageous, since collimators can be kept in the cryogenic region and therefore take advantage of the very large pumping speeds afforded by cryopumping.  Hutzler et al.\cite{Hutzler2011} found that the beam properties were largely insensitive to the placement of a collimator when using a neon buffer gas; however, as with the charcoal cryopumps, the placement of the collimator has often been seen to strongly influence the beam properties of the helium buffer gas cooled beam.  Correct placement requires careful consideration, and may involve multiple attempts.

\section{Applications}\label{sec:applications}

In this section, we will discuss some select applications where buffer gas beams can offer significant advantages: Laser cooling, precision measurements, collisional studies, and trap loading.

\subsection{Laser cooling of buffer gas beams}\label{sec:lasercooling}

Laser cooling works by continuously scattering photons off of an atom or molecule in a manner that dissipates energy\cite{Metcalf1999}. The number of photon scattering events to stop an atom or a molecule is given by $N=mv/\hbar k$, where $m$ is the mass of the species, $v$ is its velocity, and $k=2\pi/\lambda$ where $\lambda$ is the wavelength of the laser light.  For typical atoms or molecules moving at room temperature thermal velocities, this number is about $10^4$. The electronic ground state of an atom may contain multiple hyperfine states, and off-resonance excitation sometimes brings the atoms to different hyperfine states in the excited state, which can decay to an off-resonant (or ``dark'') hyperfine state in the electronic ground state.  If this happens, a ``re-pump'' laser, resonant with the other hyperfine state, can pump the atom back into the resonant state.

In contrast, the electronic ground state of molecules contains multiple internal states, including vibrational and rotational, as well as hyperfine states.
The number of re-pump lasers needed to close the optical transition grows accordingly.
Despite these challenges, Di Rosa \cite{DiRosa2004} pointed out that a set of molecules with strong single photon transitions and a highly diagonal vibrational transitions (Frank-Condon factors\cite{Herzberg1989}) may be feasible for direct laser cooling. The first criterion ensures a large photon scattering rate, meaning that a sufficient number of photons can be scattered in the limited interaction time (determined by the molecular beam velocity and geometrical constraints on the length of the interaction region).  The second criterion implies a high probability of decaying back to the initial vibrational ground state ($v''=0$). If $p\ll 1$ is the probability of decaying into a dark state after one optical transition, then, if there are no re-pump lasers, the total number of photons scattered for each molecule, $N_{scat}$, is given by $(1-p)^{N_{scat}} \sim e^{-1}$, yielding $N_{scat} \sim 1/p$.
Di Rosa\cite{DiRosa2004} listed a few molecules that satisfied the above requirements, including CaH and NH. Most of these molecules can scatter $\sim 100$ photons without any vibrational re-pump lasers, and can scatter up to $\sim 10^4$ photons with one $v''=1$ re-pump laser.

In contrast to vibrational leakage, where no strict selection rules apply, rotational transitions satisfy the total angular momentum change selection rule, $\Delta J=0, \pm 1$.
Stuhl et al. \cite{Stuhl2008} noticed that the rotational leakage can be mitigated by choosing a transition where the angular momentum $J''$ of the ground state is larger than the angular momentum $J'$ of the lowest rotational state in the excited electronic state.
For example, consider a molecule with a ground electronic state containing rotational levels $J''=1,2,3,\ldots$, and an excited electronic state with rotational levels $J'=0,1,2,\ldots$ (for example, the $X^3\Delta_1\rightarrow E^3\Pi_0$ transition of TiO\cite{Stuhl2008}).  If a laser drives the transition $|X,J''=1\rangle\rightarrow |E,J'=0\rangle$, then the selection rule $|J''-J'|\leq 1$ means that the only radiative decay allowed is $|E,J'=0\rangle\rightsquigarrow |X,J''=1\rangle$, i.e. a rotationally closed transition.  The molecule may still have vibrational leakage, but those channels can be closed with re-pump lasers in the usual way.

Several of the diatomic molecules that satisfy the criteria for laser cooling as suggested by Di Rosa\cite{DiRosa2004} and Stuhl et al.\cite{Stuhl2008} are radicals.
Since the number of photons and the distance needed to stop molecules are proportional to  $v_\|$ and $v_\|^2$ respectively, where $v_\|$ is the forward velocity of the molecules, molecular radical sources with a low forward velocity, as well as a high brightness, are desirable for direct laser cooling.  As mentioned earlier (see Table \ref{table:beamcomparison}), a buffer gas cooled beam can deliver higher brightness and lower forward velocity for these types of species.  As an example, a boosted buffer gas beam has a typical forward velocity of $\sim 150$ m s$^{-1}$, which reduces the number of photons needed to stop a CaH molecule by a factor of $\sim 5$ compared to a 1300 K oven source, as initially proposed by Di Rosa\cite{DiRosa2004}.

The first demonstration of optical cycling in molecules was achieved by using a buffer gas cooled beam of SrF\cite{Shuman2009}.  The $X^2\Sigma^+_{1/2}\rightarrow A^2 \Pi_{1/2}$ transition of SrF has relatively diagonal Franck-Condon factors\cite{Shuman2009} $f_{v'=0\rightsquigarrow v''}$: $f_{0,0}\approx 0.98$, $f_{0,1}\approx 0.02$, $f_{0,2}\approx 4\times10^{-4}$, and $f_{0,3^+}<10^{-5}$.
Theoretically, SrF could scatter $N_{scat}=1/f_{0,3^+}>10^5$ photons with two vibrational re-pump lasers, assuming that other internal states are addressed as well.
Rotational leaking was prevented by using an $N''=1 \rightarrow N'=0$ transition.
Since SrF has a nuclear spin of $I=1/2$ and the hyperfine splitting of $|N'=0,J'=1/2>$ is small, every hyperfine state in the $N''=1$ manifold would be populated during the transition and needs to be addressed.  This was achieved using lasers with several sidebands (created using an electro-optic modulator) which were on resonance with the hyperfine states.
One important feature of a transition with $J''>J'$ is that the ground state contains more Zeeman states than the excited state, leading to dark Zeeman states\cite{Stuhl2008} $m_{J''}=\pm J''$. To address these dark states, Shuman et al. used a small magnetic field which was at an angle relative to the polarization of the cooling lasers to constantly remix the Zeeman states.

With the beam source used in those first laser cooling experiments, the SrF molecules were produced by laser ablation in a helium buffer gas cell at 4 K, resulting in a molecular beam with a total number of 10$^9$ SrF molecules exiting the cell (per pulse) in the $N''=1$ state moving at $v_\|=200$ m s$^{-1}$ \cite{Shuman2009}. The molecular beam was collimated transversely by a 6 mm diameter aperture before entering the laser interaction area, which consisted of the main cooling laser, one re-pump laser for $v''=1$, and magnetic field coils. When the magnetic field was turned on in addition to the main cooling laser, Shuman et al. observed that the induced fluorescence signals increased by a factor of $\sim 3.5$. Adding the vibrational re-pump laser further increased the fluorescence signals by another factor of $\sim 3.5$, corresponding to a total average number of photons scattered per molecule of $N_{scat}\sim 170$. Since both lasers were applied perpendicularly to the molecular beam from one transverse direction, the net momentum transferred from the photons to SrF was sufficient to create a position shift or deflection of the beam. The observed deflection indicated $N_{scat}= 140$, consistent with the previous estimation based on the fluorescence measurement.

In a subsequent experiment, Shuman et al. \cite{Shuman2010} demonstrated transverse cooling of the buffer gas cooled SrF beam.
By incorporating an additional re-pump laser for the $v''=2$ state and arranging the cooling lasers to intersect the molecular beam multiple times at nearly 90$^{\circ}$, the molecules scattered $N_{scat}=500-1000$ photons in a 15 cm long region. When the cooling lasers were red detuned with a magnetic field of 5 G, the transverse temperature of the SrF beam was cooled from $T_0=50$ mK (due to the geometry of the mechanical beam collimators) down to $T\sim 5$ mK. In this regime, the mixing field is large enough for rapidly remixing molecules in the Zeeman states, leading to Doppler cooling. Shuman et al. discovered that the molecular beam also obtained a lower temperature when the lasers were blue detuned with a relatively low field of 0.6 G. In this case, SrF in the bright Zeeman state constantly experienced a periodic potential provided by the cooling lasers, resulting in ``Sisyphus''-like cooling\cite{Metcalf1999,Shuman2010}.

\subsubsection{Direct loading of exotic atoms into magneto-optical traps}\label{sec:motloading}

Several exotic atoms have been successfully loaded into magneto-optical traps (MOTs\cite{Raab1987,Metcalf1999}), including Yb \cite{Takahashi2004,Pandey2010}, Ag\cite{Uhlenberg2000}, Er\cite{McClelland2006}, Dy \cite{Youn2010} and Tm \cite{Sukachev2010}.
Because these atoms have much higher melting points than alkali atoms, effusive oven sources used to create these atomic beams operated at high temperatures, between 700 and 1500 K.
A Zeeman slower is typically needed to decelerate the fast-moving atoms ($\bar{v}_\| \sim 350-580$ m s$^{-1}$) for efficiently loading into MOT, which has a typical capture velocity $< 60$ m s$^{-1}$.
As discussed in Section \ref{sec:slowingcells}, a near effusive buffer gas cooled beam can be produced by adding one slowing cell to the typical buffer gas cell. Although this type of beam has been found to have a typically lower extraction efficiency$(\sim 1 \%)$ than a hydrodynamically extracted beam without a slowing cell, the low forward velocity of the beam makes direct loading of exotic atomic beams into MOTs feasible.  Additionally, the very high fluxes possible with ablation of certain metals can lead to a superior beam system.

The longitudinal and transverse temperatures of CaH molecules in a beam produced with a slowing cell were measured to be $\sim 5$ K and $\sim 2.5$ K, respectively \cite{Lu2011}. Since the cooling mechanism only relies on elastic collisions with the He buffer gas, we expect atomic beams produced from such a slowing cell to have properties similar to the beam of CaH (except, of course, for any possible mass-related differences).
For example, the forward velocity of a slow Yb beam with a longitudinal temperature of 5 K is $\sim 30$ m s$^{-1}$, which is consistent with the forward velocity of a Yb beam demonstrated by Patterson \cite{Patterson2010} using a similar cell design.
Employing a one-dimensional laser cooling model, the capture velocity of a MOT is given by $v_c=(|\Delta|+\Gamma)/k$ \cite{Pandey2010}, where $\Delta$ is the detuning of the cooling laser and $\Gamma$ is the natural linewidth. For Yb, assuming $|\Delta|=\Gamma=2\pi\times 28$ MHz, the capture velocity is 22 m s$^{-1}$, implying a large fraction of a buffer gas cooled Yb beam can be loaded without any longitudinal laser slowing (see Table \ref{table:MOTefficiency}).

Collisional heating from the buffer gas streaming through the MOT region could be a concern. Typically, the buffer density inside the slowing cell is on the order of $10^{14}$ cm$^{-3}$. Although it is difficult to measure the He beam profile, a good approximation is to assume that the He beam attains the same temperatures as the CaH molecular beam.
Therefore, the collisional heating from the buffer gas is negligible compared to the cooling power from the scattered photons starting from a distance of 10 cm away from the aperture, where the He density is $<10^{12}$ cm$^{-3}$.
Any potential loss due to the divergence of the beam can be mitigated by employing transverse laser cooling to collimate the atomic beam between the cell aperture and MOT.
Assuming the transverse capture velocity is given by $v_{c{,t}}=\Gamma/k$, we can estimate the loading efficiency of a buffer gas cooled beam into MOTs by considering the velocity distributions that are within $v_{c{,t}}$ transversely and $v_c$ longitudinally.

Table \ref{table:MOTefficiency} shows some exotic atoms that have been loaded into MOTs from effusive oven sources, and have also been buffer gas cooled. Of these species, only Yb has been realized as a buffer gas beam so far\cite{Patterson2007}.  Given the generality of the buffer gas beam technique, we believe that producing cold and slow exotic atomic beams with beam properties similar to the CaH beam\cite{Lu2011} is very likely.
We can therefore estimate the loading efficiency of these atoms into MOTs from buffer gas beams, $f_{load}$, by assuming that these proposed atomic beams would have the same longitudinal and transverse temperatures of the CaH beam, 5 K and 2.5 K respectively.
Based on the ablation yields inside the buffer gas cell, $N_{ablation}$, reported in the past\cite{Campbell2009} and a typical extraction efficiency of the slowing cell of $f_{ext}\sim 5\times 10^{-3}$, we can estimate potential loaded numbers per ablation pulse $N_{load}\sim N_{ablation}\times f_{ext}\times f_{load}$, shown as the last column in Table \ref{table:MOTefficiency}.
The estimated numbers of atoms loaded per pulse are high compared to the reported trapped numbers. 
By loading $1-10$ pulses, the expected atom numbers (or densities) accumulated in MOTs could approach a regime where the radiation pressure and collisions between ground and excited state atoms would limit the atomic density\cite{Ketterle1993}.
Co-loading multiple species without using high temperature ovens and Zeeman slowers into a MOT is feasible by simply ablating several species simultaneously.

\begin{table}
	\centering
		\begin{tabular}{llllll}
			Atom &$N_{trapped,exp}$ & $f_{load}$ [$\%$] & $N_{load}$ [pulse$^{-1}$] \\ \hline
			Yb \cite{Pandey2010} &  $10^7$           &21   & $2\times 10^{10}$ \\
			Ag \cite{Uhlenberg2000}&  $3\times10^6$  &1.7  & $3\times 10^{9}$ \\
			Dy  \cite{Youn2010}&  $2.5\times 10^8$   &33   & $3\times 10^{9}$ \\
			Er  \cite{McClelland2006} &  $10^6$      &39   & $4\times 10^{8}$ \\
			Tm \cite{Sukachev2010}&  $7\times10^4$   &0.3  & $3\times 10^{6}$ \\
		\end{tabular}
	\caption{Some exotic atoms which have been loaded into a MOT from an effusive oven source, and also buffer gas cooled.  $N_{trapped,exp}$ is the number of atoms that were loaded into the MOT in the respective experiments.  $f_{load}$ is the projected loading efficiency from a buffer gas beam source, and $N_{load}$ is the projected number of atoms which could be trapped in the MOT with a single pulse from a buffer gas source.  A description of these estimates may be found in the text.}
	\label{table:MOTefficiency}
\end{table}

\subsection{Precision measurements}\label{sec:precisionmeasurements}

Another application of molecular beams is the precision determination of molecular properties.  This is made possible by the lack of collisions in the beam region, and enhanced by the large volume of advanced molecular beam production, manipulation, and spectroscopy techniques\cite{Ramsey1985,Scoles1988}.  Precision spectroscopy with supersonic beams is a workhorse in chemistry and physics.  As the reader can surely surmise by now, many precision measurements can benefit from the high flux and low forward velocity afforded by buffer gas cooled beams.  In this section we will review some molecular beam precision measurements, and discuss how they could benefit from buffer gas cooled beam technology.

\subsubsection{Electric dipole moments}

In 1950, Purcell and Ramsey\cite{Purcell1950} were the first to seriously consider the possibility of a fundamental particle (i.e. an electron or nucleon) possessing a permanent electric dipole moment (EDM).  Before this time, it was widely (though not universally\cite{Dirac1949}) believed that a fundamental particle could not possess such a moment because it would violate parity (P) and time-reversal (T) invariance in the following manner: Since a dipole moment is a vector quantity it must be aligned with the spin of the particle, which is the only intrinsic vector associated with a fundamental particle.  However, an electric dipole moment is a polar vector, while spin is an axial (or pseudo-) vector, so these two quantities must behave oppositely under inversion of coordinate systems or reversal of time, yet the previous argument requires them to uniquely determine each other.  Although no permanent EDM of a fundamental particle has ever been measured (despite over 50 years of experimental searches\cite{Smith1957}), such a discovery would have a profound impact on our understanding of particle physics and cosmology.  For an accessible discussion about EDMs of fundamental particles and their implications, see the article by Fortson, Sandars, and Barr\cite{Fortson2003}; for more details, see the book by Khriplovich and Lamoreaux\cite{Khriplovich1997}.

The essential idea of nearly every EDM search is to look for a small shift in precession frequency of a spin in an electric field as the electric field direction is reversed\cite{Khriplovich1997}.  For a particle with spin $\vec{S}$, magnetic dipole moment $\mu$ and electric dipole moment $d$ in a parallel magnetic field $\vec{\mathcal{B}}$ and electric field $\vec{\mathcal{E}}$, the spin of the particle (which is parallel to both the magnetic and electric dipole moment) will precess about the fields with angular frequency
\be \omega = \frac{d\vec{\mathcal{E}}+\mu\vec{\mathcal{B}}}{\hbar}\cdot\frac{\vec{S}}{|\vec{S}|}.\ee
Consider an experiment where $\vec{S}$ and $\vec{\mathcal{B}}$ are kept fixed but the direction of $\vec{\mathcal{E}}$ is reversed, and then the frequency $\omega_\pm$ is measured for the two different orientations $\pm\vec{\mathcal{E}}$.  The value of the electric dipole moment $d$ can then be extracted by comparing these two measured values,
\be d = \frac{\hbar(\omega_+-\omega_-)}{2\vec{\mathcal{E}}\cdot\vec{S}/|\vec{S}|}.\ee
The determination of $d$ then reduces to a precision measurement of the frequencies $\omega_\pm$.  Precise frequency measurements are best accomplished by the separated oscillatory field (or Ramsey) method\cite{Ramsey1985}, in which the spins are allowed to precess for a time (the ``coherence time'') $\tau$ before the accumulated phase angle is measured, thereby determining $\omega$.  If such an experiment is repeated with a total of $N$ spins, the uncertainty in the EDM is given by\cite{Khriplovich1997}
\be \delta d = \frac{\hbar/\tau}{|\vec{\mathcal{E}}|\sqrt{N}/|\vec{S}|}. \label{dedm}\ee

From this equation, we can see that an EDM experiment should have large electric fields, long coherence times, and large count rates.  Neutral objects such as neutrons or atoms can simply be placed in large electric fields; however; charged particles like electrons or protons would simply accelerate out of the electric field.  Sandars\cite{Sandars1965} realized that not only could an electric dipole moment of a charged object be inferred by studying a composite neutral object, such as an atom or molecule, but the electric dipole moment could actually be enhanced.  For example, Sandars calculated\cite{Sandars1966} that an electric dipole moment of the electron $d_e$ could induce an electric dipole moment of an atom $d_A$, with a relativistic ``enhancement factor'' $d_A/d_e$ that can be over 100 for Cs atoms.  The basic idea\cite{Commins2007} is that the electron can sample the atomic electric field, which is typically of order
\be \mathcal{E}_{atomic}\approx \frac{e^2}{4\pi\epsilon_0a_0^2}\approx 5 \textrm{ GV cm}^{-1},\ee
 where $a_0$ is the Bohr radius.  This is considerably larger than a typical macroscopic lab field of $\lesssim 100$ kV cm$^{-1}$.  Atoms cannot take full advantage of this enhancement because they cannot be fully polarized in laboratory electric fields\cite{Commins2010}; however, polar molecules can be polarized in lab fields, and for this reason most current electron EDM experiments use polar molecules\cite{Commins2010,Commins2010}.

The molecules which have the highest enhancement factor have high-$Z$ atoms with partially filled electron shells, and tend to be chemically reactive, have high melting points, and are often free radicals\cite{Commins2010,Commins2010}.  For this reason, effusive beam sources of these molecules are typically not feasible; the Boltzmann distribution of rotational states at temperatures where the species have appreciable vapor pressure put only a tiny fraction in any single rotational state (however, performing an electron EDM experiment with molecules in a hot vapor cell is a possibility\cite{Bickman2009}).  A molecular beam for an EDM search should therefore have low temperature and, in light of Eq. (\ref{dedm}), a low forward velocity to allow the molecules to interact with the fields for a long time, giving large $\tau$.  For the types of molecules under consideration, buffer gas cooled beams\cite{Maxwell2005,Hutzler2011,Barry2011} have considerable advantages over the standard supersonic beam\cite{Tarbutt2002}.  As discussed earlier, the brightness for refractory or chemically reactive species can be very high with buffer gas beams, and they have considerably slower forward velocities.  In Table \ref{table:beamcomparison}, we can see that a buffer gas cooled ThO beam has a slower forward velocity and over 100 times the brightness compared to a demonstrated, state-of-the-art YbF supersonic beam, both of which are currently being used for an electron EDM search.  These reasons make buffer gas cooled beams an attractive option for electron EDM searches, with some existing supersonic experiments considering switching to buffer gas cooling\cite{Skoff2011}.

Searches for the electron EDM are currently underway in molecular beams of YbF\cite{Hudson2011}, ThO\cite{Vutha2010}, and WC\cite{Lee2009}, as well as several other atomic, molecular, and solid state systems\cite{Commins2010}.

\subsubsection{Parity violation and anapole moments}

The exchange of weak neutral currents between the constituent particles of an atom or molecule can lead to observable parity violating effects in their spectra.  Precision examinations of these effects, known as atomic parity violation (APV), allow the study of nuclear and electroweak physics using atomic systems\cite{Ginges2004}.  An experimentally feasible approach to using APV to probe electroweak physics was first proposed in 1974,\cite{Bouchiat1974} and since then APV has been observed in a number of atoms\cite{Guena2005,Ginges2004,Tsigutkin2009,Barkov1978}.  Here we will briefly discuss how buffer gas beams may improve experimental studies of APV; detailed discussions of APV theory and experiments may be found elsewhere\cite{Ginges2004,Guena2005,Dzuba2010,Budker2008}.

\emph{Weak neutral currents.}  The exchange of a neutral $Z_0$ boson between the nucleus and the electrons in an atom can give rise a parity-violating, electronic contact potential\cite{Bouchiat1974}. This potential can mix atomic states of opposite parity, such as a valence electron in $S_{1/2}$ and $P_{1/2}$ atomic orbitals.  This mixing can either be measured by searching for parity violating optical rotation in atomic emission/absorption\cite{Barkov1978,Bouchiat1974}, or by looking for interference between the APV induced state mixing and the mixing provided by an external electric field (the Stark-interference method\cite{Conti1979}).  APV has been measured in a number of atoms\cite{Ginges2004,Tsigutkin2009}, and comparison to theory gives tests of the standard model at low energies.

\emph{Nuclear anapole moments.}  Another parity violating effect in an atom is the anapole moment\cite{Zel'dovich1957,Ginges2004,Budker2008}, which arises from an exchange of a $Z_0$ or $W_{\pm}$ boson between nucleons in a nucleus.  Like the parity-violating potential above, the interaction is a contact potential between the electrons and nuclear anapole moment.  The observable effect is very similar to that of the weak neutral currents discussed above, with the important exception that the size of the effect will vary depending on the nuclear spin.  An experimental signature of an anapole moment is then a difference in parity violating amplitudes between different hyperfine levels.  So far only a single anapole moment has been measured, in the $^{133}$Cs nucleus\cite{Wood1997}, though other experiments are underway\cite{Dzuba2010}.

\emph{Enhancement of parity violation in diatomic molecules}.  Parity-violating effects have thus far been observed only in atoms, though the search has been extended to diatomic molecules for very compelling reasons.  Molecules have states of opposite parity (rotational states, and in some cases $\Lambda/\Omega$-doublets\cite{Herzberg1989}) which are typically much smaller than the spacing between opposite parity electronic states in atoms, so the effects of nuclear-spin-dependent parity-violating violation (such as anapole moments) can be greatly enhanced\cite{Labzovskii1978,Sushkov1978,DeMille2008,Flambaum1985}.  This mixing can be increased even further by applying external magnetic fields to push rotational states to near-degeneracy.  DeMille et al.\cite{DeMille2008} proposed (and are working on) an experiment to use a supersonic beam of BaF free radicals in a $^2\Sigma$ electronic state to measure parity-violating amplitudes with high precision.

Similar to the case with an EDM measurement, the shot-noise limited uncertainty (Eq. (\ref{dedm})) in the parity violating amplitude scales as $\propto N^{-1/2}T^{-1}$, where $N$ is the total number of interrogated molecules and $T$ is the time that the molecules spend interacting with the electromagnetic fields.  Also similar to the EDM measurements is the choice of molecules, including\cite{Dzuba2010} BiO, BiS, HgF, LaO, LaS, LuO, LuS, PbF, and YbF: These are all free radicals with heavy nuclei and are therefore prime candidates for production in a buffer gas cooled beam.  There are even some molecules which have already been discussed, such as YbF and BaF (Table \ref{table:beamcomparison}), for which using a buffer gas beam can deliver over 100 times the brightness of a supersonic beam, with a slower forward velocity (allowing for the same time $T$ with a shorter apparatus, easing technical requirements).  It is for this reason that a BaF anapole moment experiment (at Yale\cite{DeMille2008}) is pursuing a buffer gas cooled beam source\cite{Rahmlow2010}.

\subsubsection{Time-variation of fundamental constants}

There has been much recent interest in the question of whether or not fundamental constants are truly ``constant,'' or whether their values have changed over time.  In particular, the dimensionless fine structure constant $\alpha=e^2/4\pi\epsilon_0\hbar c$ and electron-proton mass ratio $\mu=m_e/m_p$ have attracted special attention due to the possibility of measuring their variation from multiple independent sources.  Detailed discussions of the theory and experiments discussed here may be found in reviews\cite{Flambaum2009,Chin2009}.  

Searches for time-variation of fundamental constants typically take one of two approaches.  One is to look at data which may give information about $\alpha$ or $\mu$ from a very long time ago, for example by examining astronomical spectra at large redshifts\cite{Flambaum2009,Flambaum2007a}, but with limited accuracy (though the search may be coupled with a laboratory precision spectroscopy experiment).  Another technique is to use measurements with high precision but over a comparatively short timescale, for example by using an atomic clock\cite{Rosenband2008,Blatt2008}.  The current limits on $\mu$ and $\alpha$ are
\[ \frac{\dot{\mu}}{\mu} = (1\pm 3)\times10^{-16}\textrm{ yr}^{-1} \]
from the inversion spectrum of ammonia measured in sources with high redshift\cite{Flambaum2007a}, and
\[ \frac{\dot{\alpha}}{\alpha} = (-1.6\pm 2.3)\times10^{-17}\textrm{ yr}^{-1} \]
from comparing optical clocks\cite{Rosenband2008}.

As with EDMs and parity violation, molecules can have large enhancement factors which could allow for more sensitive measurements.  In particular, the $\alpha$ and $\mu$ dependence of fine structure splitting $\omega_f\propto\alpha^2$, rotational splitting $\omega_r\propto\mu$ and vibrational splitting $\omega_v\propto\mu^{1/2}$ are different\cite{Flambaum2009}, and if two levels with different hyperfine, rotational, or vibrational character are nearly degenerate there can be significant sensitivity enhancement\cite{Flambaum2007} in changes to both $\mu$ and $\alpha$.  This enhancement can be several orders of magnitude (compared to atoms), and there are several promising species\cite{Flambaum2009}, including CuS, IrC, LaS, LaO, LuO, SiBr, YbF, Cs$_2$, and Sr$_2$.  Several of these molecules may be difficult to produce through normal beam methods, but are prime candidates for buffer gas beam production.

\subsection{Cold collisions and trap loading}\label{sec:collisions}

The study of cold molecular collisions may impact areas of fundamental physics, astrophysics, and chemistry.
For example, several cold neutral-neutral reactions are calculated to proceed with a large rate \cite{Sabbah2007,Frankcombe2007,Edvardsson2006}, and hence may play crucial roles in cold chemistry of interstellar clouds, where low mass stars may be formed\cite{Fraser2002}.
The number of direct measurements of reaction rates involving cold molecules is small compared to the number of theoretical calculations, but new measurements could be used in detailed astrophysical modeling, and can assist further theoretical calculations\cite{Sabbah2007,Frankcombe2007,Hummon2008}.

Collision studies at high temperatures may involve many internal states and scattering channels, which can complicate both theory and measurements; cooling the sample to a few K can reduce the involved states and channels, making the problem more tractable.  In addition, the potential energy provided by electromagnetic fields on magnetic or polar molecules (typically a few Tesla or kV cm$^{-1}$, respectively) can be comparable to the thermal energy of the cold molecule, which makes the experimental control of collisions and chemical reactions an achievable goal\cite{Krems2008,Carr2009}.

Understanding cold molecular collisions is an important step for producing ultra-cold molecules via direct cooling methods.
Cold molecules produced from direct methods (buffer gas cooling, supersonic expansion with Stark deceleration, etc.) so far possess a thermal energy $>k_B\times1$ mK. There are a few possible paths to approach ultra-cold temperatures, including evaporative cooling and sympathetic cooling of molecules.
The former (latter) relies on removing molecules (atoms) from the high energy tail in a thermal distribution, followed by subsequent elastic molecule-molecule (atom-molecule) collisions which rethermalize molecules to a lower temperature.
Therefore, good molecule-molecule or atom-molecule collision properties are required for bringing cold molecules into ultra-cold temperatures.
  
Buffer gas cooling can provide a large molecular sample in a single quantum state with a kinetic energy on the order of $\sim k_B\times$ K.
Several cold atom-molecule and molecule-molecule systems have been investigated using buffer gas cooling, including collisions of He atoms with $^2\Sigma$ and $^3\Sigma$ molecules\cite{Campbell2007,Weinstein1998}, N + NH\cite{Hummon2011}, and ND$_3$ + OH\cite{Sawyer2011}.

The He + molecule and N + NH collision systems involved magnetic trapping of molecules and resulted in deeper understanding of atom-molecule interactions.
For example, the He + $^2\Sigma$ and $^3\Sigma$ molecule collision systems revealed the spin relaxation mechanisms of $^2\Sigma$ and $^3\Sigma$ molecules in magnetic traps\cite{Campbell2009}.
Cold N + NH collisions indicated not only that N might be a good candidate for sympathetic cooling of NH, but that a cold chemical reaction N + NH $\rightarrow$ N$_2$ + H may be suppressed in the magnetic field\cite{Hummon2011}.

\subsubsection{ND$_3$ + OH collisions}
Trapping molecules provides an ideal environment for studying cold molecular collisions by providing a long interaction time.
Using trapped molecules as the scattering target and monitoring the molecular trap loss allows absolute measurements of the collision cross sections \cite{Sawyer2008}.
Although several chemically interesting molecules have been cooled via direct methods and trapped, inelastic molecular collisions or reactions have not been observed in the trap because of a low molecular density and a low collisional energy ($\sim 100$ mK).
Assuming a typical elastic cross section of $10^{-14}$cm$^2$, an elastic NH + NH collision occurs in hundreds of seconds with a molecular density of $10^8$ cm$^{-3}$ at 0.5 K, which is much longer than the trap lifetime of $\sim$1 s in the buffer gas loading experiment\cite{Campbell2007}.

Absolute atom-molecule cross section measurements were obtained for the OH+D$_2$ and OH+He systems by scattering supersonic D$_2$ and He beams off of magnetically trapped OH molecules\cite{Sawyer2008}.
The lowest collisional energies (84 K for He and 200 K for D$_2$) achieved with a cooled supersonic nozzle were two order of magnitudes higher than the trapping potential of the molecules. In such a situation it is difficult to differentiate the contributions of elastic and inelastic collisions to the total cross section and control the collisions with an external field.
Since buffer gas cooled beams of molecules can be cold, slow, and bright, they are an attractive choice for the study of molecular collisions.
In addition, several buffer gas beams can provide continuous output (see Table \ref{table:beamproperties}), which could allow molecular interaction times comparable to the lifetime of the trapped species.

Such an experiment was performed by Sawyer et al. \cite{Sawyer2011} by combining a buffer gas cooled beam of ND$_3$ with trapped OH molecules.  Collisions were studied on the $\sim$1 s timescale, and led to the first observation of cold (5 K) heteromolecular dipolar collisions.
The continuous ND$_3$ beam was guided to suppress the background buffer gas, giving a typical molecular density of $10^8$ cm$^{-3}$ and a mean velocity of 100 m s$^{-1}$ before colliding with OH.
A supersonic OH beam was Stark-decelerated and loaded to a magnetic trap with a density of $\sim 10^6$ cm$^{-3}$ at a temperature of 70 mK.
To study the ND$_3$ + OH collisions, the trap lifetime of OH in the presence of an ND$_3$ beam was compared to the lifetime solely limited by collisions with the background gas, yielding a trap loss cross section of $\sigma_{{exp}}^{{loss}}=0.27\times 10^{-12}$ cm$^2$.
At a collision energy of few K, only tens of partial waves participate in the collision, and hence external fields can affect the collisional properties.
After applying a large electric field to polarize both OH and ND$_3$, $\sigma_{{exp}}^{{loss}}$ increased by a factor of 1.4.

Sawyer et al. also performed both quantum scattering and semi-classical calculations to compute the elastic and inelastic OH + ND$_3$ cross sections.
The calculated cross sections were orders of magnitude larger than typical gas-kinetic values, possibly resulting from the dipolar interaction.
To calculate the cross sections in the field, Sawyer et al. included a long range and anisotropic dipolar interaction term $C_3/r^3$, in addition to an isotropic van der Waals interaction for the zero field case.
They discovered that the elastic cross section increased by a factor of 2.7 in the presence of a field, compared to a relatively constant inelastic cross section.
After considering the probability of an elastic scattering leaving OH un-trapped, Sawyer et al. found the value of $\sigma_{{exp}}^{{loss}}$ and its increase in the field were consistent the theoretical model, indicating the elastic OH + ND$_3$ collisions not only dominate the trap loss but can be controlled by the external electric field.

Compared to other beam technologies, buffer gas cooled beams are well suited for these types of collision experiments.  Stark-deceleration of a supersonic ND$_3$ source can yield a beam with a forward velocity of $\sim 90$ m s$^{-1}$ and a peak density of $\sim 10^8$ cm$^{-3}$, but with pulsed operation, $<1$ ms pulse duration, and repetition frequency of few Hz\cite{Bethlem2002}.
The fractional loss of trapped molecules due to colliding with another beam is given by
\begin{equation} \Delta N/N=-k^{{loss}} \Delta t=-n_{{beam}}\sigma^{{loss}}v_{{rel}}\Delta t, \end{equation}
where $n_{{beam}}$ is the density of the beam and $v_{{rel}}$ is the mean relative velocity.
Using a continuous buffer gas cooled ND$_3$ beam increases the average interaction time by a factor of $\sim$100 compared to a Stark-decelerated ND$_3$ beam, leading to a higher sensitivity in observing cold collisions.

\subsubsection{Direct trap loading of molecules}\label{sec:traploading}
Cold and slow beams of CaH, with a kinetic energy of few $k_b\times$K, can be created by using a slowing cell (see Section \ref{sec:slowingcells}), which should allow direct loading of the molecules into a magnetic trap.  A magnetic lens or guide can be used to ensure that the buffer gas atoms do not interfere with the trapped molecules, similar to the experiment of Sawyer et al.\cite{Sawyer2011} discussed earlier.
Here we will explore one possible loading scheme, which combines magnetic slowing and optical pumping to achieve trap loading of slow CaH molecules.  By using a buffer gas cooled beam, the proposed method benefits from large molecular fluxes, chemical versatility, and high vacuum in the trapping area.  A similar scheme had been employed to continuously load an atomic Cr beam into an optical dipole trap\cite{Falkenau2011}.  

Figure \ref{fig:loadingscheme} illustrates the scheme to continuously load CaH molecules into a quadrupole trap.
When a low-field seeking molecule ($m_j^{\prime\prime}=1/2$) enters the trap, it travels up the potential hill and slows down.
Near the saddle point, a laser can optically pump the molecules to the high-field seeking state ($m_j^{\prime\prime}=-1/2$).
Molecules continue to be decelerated while approaching the center of the trap, when another laser pumps the molecule into the trapped $m_j^{\prime\prime}=1/2$ state.
In this loading scheme, molecules with a kinetic energy large enough to overcome two potential hills can reach the trap center.
Molecules with a kinetic energy lower than the trap depth after deceleration will remain trapped.
For a magnetic quadrupole trap with a depth of 4 T, the capture energy for CaH is 5--7 K, suggesting that a large fraction of the slow CaH beam can be captured.

To reduce the background buffer gas density in trap area, a magnetic lens can be used to collimate the molecular beam while the buffer gas is pumped away with charcoal cryopumps (see Section \ref{sec:technicaldetails}).
We performed a semi-classical Monte Carlo simulation to calculate the efficiency of loading the CaH beam into the 4 T deep quadrupole trap, indicating a typical efficiency of $\sim 5\times 10^{-3}$, or about $3\times 10^6$ trapped molecules per pulse. The background gas density in the trapping region can be kept low such that molecules can be accumulated in the trap via continuously loading, leading to a total number of $10^7-10^8$ over time. This compares favorably to the typical values of $10^{4-5}$ molecules loaded into a trap from a Stark-decelerated supersonic source\cite{Meerakker2005,Meerakker2005II,Bethlem2000,Gilijamse2007,Hoekstra2007}.

We would like to point out that this scheme is general and applicable to other magnetic species including atoms.
More importantly, it only requires scattering a few photons to achieve loading and therefore does not require a highly closed cycling transition.  This opens up the possibility of trapping a large class of magnetic molecules.

\begin{figure}[htbp]
  \includegraphics[width=3in]{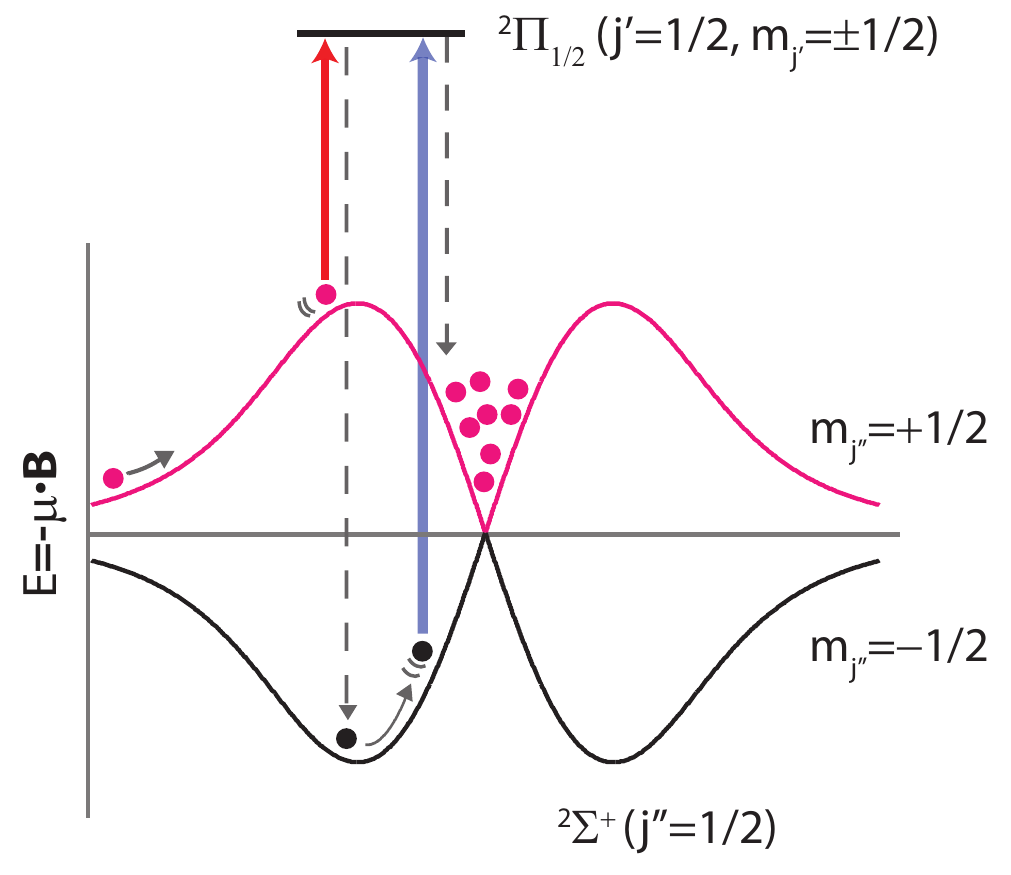}
  \caption{Loading scheme of a slow buffer gas beam into a magnetic trap. Pink and black curves indicate the potential energy of the low-field and high-field seeking states in a magnetic trap, respectively. Red and blue arrows show the optical pumping processes to achieve irreversible loading due to spontaneous emissions.}
  \label{fig:loadingscheme}
\end{figure}

\section{Acknowledgments}

Thanks to John Barry, Sid Cahn, Dave DeMille, Richard Hedricks, Emil Kirilov, Dave Patterson, Dave Rahmlow, Julia Rasmussen, and Joe Smallman for helping us gather information for this paper, and to Colin Connolly and Elizabeth Petrik for feedback on the manuscript.

\bibliography{BufferBeamChemRev2011bib}
\end{document}